\input amstex
\documentstyle{amsppt}
\magnification=\magstep1
\NoBlackBoxes
\def\today{\ifcase\month\or
 January\or February\or March\or April\or May\or June\or
 July\or August\or September\or October\or November\or December\fi
 \space\number\day, \number\year}

\baselineskip=32pt
\parindent=18pt
\def\nind{\noindent}
\def\RR{\text{{\rm I \hskip -5.75pt R}}}
\def\IN{\text{{\rm I \hskip -5.75pt N}}}

\def\CC{\;\text{{\rm \vrule height 6pt width 1pt \hskip -4.5pt C}}}

\def\As{{\Cal A}}
\def\Bs{{\Cal B}}

\def\Fs{{\Cal F}}
\def\Gs{{\Cal G}}
\def\Hs{{\Cal H}}

\def\Ks{{\Cal K}}
\def\Ls{{\Cal L}}

\def\Us{{\Cal U}}

\def\nphi{{n_{\varphi}}}

\def\idty{{\leavevmode\hbox{\rm 1\kern -.3em I}}}
\def\da{{\dagger}}

\def\FH{{\Cal F (\Cal H)}}
\def\Fo{{\Cal F _0}}
\def\L2Q{{L^2(Q,d\mu)}}
\def\LiQ{{L^{\infty}(Q,d\mu)}}
\baselineskip 15pt plus 2pt
\spaceskip=.5em plus .25em minus .20em
\redefine\qed{\hbox{$\boxed{}$}}

\topmatter
\title Further Representations of the Canonical Commutation Relations   
\endtitle

\author   Martin Florig and Stephen J. Summers 
\endauthor     

\address{Department of Mathematics, University of Florida, Gainesville, FL  
32611, USA } \endaddress

\date{August 1997, revised December 1998} \enddate

\abstract{We construct a new class of representations of the canonical
commutation relations, which generalizes previously known classes. We
perturb the infinitesimal generator of the initial Fock representation 
({\it i.e.} the free quantum field) by a function of the field which is
square-integrable with respect to the associated Gaussian measure. We
characterize which such perturbations lead to representations of the canonical
commutation relations. We provide
conditions entailing the irreducibility of such representations, show 
explicitly that our class of representations subsumes previously studied 
classes, and give necessary and sufficient conditions for our representations
to be unitarily equivalent, resp. quasi-equivalent, with Fock, coherent or
quasifree representations.}
\endabstract
\endtopmatter

\newpage

\document

\heading I. Introduction  \endheading

     The canonical commutation relations (henceforth the CCR) were initially
introduced in 1927 by Dirac as generalizations of Heisenberg's commutation
relations in order to discuss radiation theory \cite{10}. Since then, the
CCR have proven to be of central importance in the quantum description of
bosonic systems \cite{6}, {\it i.e.} systems of identical 
particles satisfying Bose-Einstein statistics, such as the photons Dirac 
considered. The most elementary formulation of the CCR for $n$ degrees of
freedom is
$$[q_j,p_k] \equiv q_j p_k - p_k q_j = i\delta_{jk} \cdot \idty \quad , 
\quad j,k \in \{1,\ldots,n\} \quad , $$
where $\delta_{jk}$ is the Kronecker delta, $\idty$ is the identity
operator on the complex Hilbert space $\Ks$, and $q_j,p_k$ are symmetric 
operators on $\Ks$. It is well-known that the operators $q_j,p_k$ must be 
unbounded, and so the CCR are to be understood on a suitable dense subspace of 
$\Ks$. Under certain circumstances, these operators may be exponentiated
to obtain unitaries 
$$U_k(a) = e^{-iap_k} \quad \text{and} \quad V_j(a) = e^{-iaq_j} \quad , $$
with $a \in \RR$, satisfying
$$\align
U_k(a)V_j(b) &= e^{iab}V_j(b)U_k(a) \quad , \quad j = k , \\
U_k(a)V_j(b) &= V_j(b)U_k(a) \quad , \quad j \neq k , 
\endalign $$
for all $a,b \in \RR$, $j,k = 1,\ldots,n$. These relations constitute the Weyl 
form of the 
CCR for $n$ degrees of freedom. These unitaries generate a $C^*$-algebra on 
$\Ks$. A representation of (the Weyl form of) the CCR is a $C^*$-homomorphism 
preserving the Weyl relations. A remarkable early result about the CCR is the
Stone-von Neumann uniqueness theorem \cite{20}: all irreducible 
representations of the CCR for $n$ degrees of freedom are unitarily equivalent.
\par
     When rigorous mathematical analysis of the CCR in the case of infinitely 
many degrees of freedom --- the case of relevance to quantum statistical 
mechanics and quantum field theory --- began in the 1950's, it was quickly 
realized that there are uncountably infinitely many unitarily inequivalent 
irreducible representations of the CCR in this case \cite{12} and that the 
choice of proper 
representation is crucial in any physical application. This last point deserves
emphasis. It has become clear from rigorous study of concrete models in
constructive quantum field theory that bosonic systems with identical 
kinematics but physically distinct dynamics require inequivalent 
representations of the CCR. Roughly speaking, the kinematical aspects 
determine the choice of CCR-algebra, whereas the dynamics fix the choice of 
the representation of the given CCR-algebra in which to make the relevant, 
perturbation-free computations. (It is also believed --- and proven in a 
number of indicative special cases --- that perturbation series in one 
representation provide divergent and at best asymptotic approximations to the 
physically relevant quantities in another, unitarily inequivalent 
representation.) For an overview of these matters and further references,
the reader is referred to \cite{37}. \par
     Many classes of representations of the CCR for infinitely many degrees of
freedom have been rigorously studied in the literature - we mention, in 
particular, the Fock \cite{7}, coherent \cite{31}, quasifree \cite{27} (or 
symplectic), infinite product \cite{15}, and, more recently, quadratic 
\cite{26}\cite{22}\cite{29} and polynomial representations \cite{30}.
In this paper, we wish to construct and prove certain mathematically and
physically relevant properties of a broad class of representations of the CCR 
containing all of the above (except possibly the infinite product 
representations) as special cases. \par
     We suggest heuristically the nature of these representations with the
following easily describable examples. Let $\{q_k,p_k\}_{k=1}^{\infty}$ be a 
system of densely defined operators acting on the complex Hilbert space $\Ks$ 
and satisfying
$$[q_j,p_k] = i \delta_{jk} \cdot \idty $$
on a suitable dense subset of $\Ks$, and for the standard annihilation and 
creation operators $a_k \equiv \frac{1}{\sqrt{2}}(q_k + ip_k)$ and 
$a_k^{\dagger} \equiv \frac{1}{\sqrt{2}}(q_k - ip_k)$ let there be a vector
$\Omega \in \Ks$ such that $a_k\Omega = 0$ for all $k \in \IN$. In other 
words, we consider the Fock representation\footnote{Often called the Fock-Cook 
representation, since it was in \cite{7} that this representation was given a
mathematically rigorous and relativistically covariant form.} of a bosonic 
system with infinitely many degrees of freedom \footnote{in the so-called 
basis-dependent formulation --- We discuss the basis-independent formulation 
and discuss their relation in Chapter II.} (see \cite{7}\cite{6}).
In this setting, the coherent, resp. quasifree, quadratic or polynomial 
canonical transformations can be written as 

$$\align
  (coherent) \quad q_k \mapsto q_k, \qquad &p_k \mapsto p_k + \lambda_k \cdot \idty \quad , \\
  (quasifree) \quad q_k \mapsto q_k,  \qquad &p_k \mapsto p_k + 
                              \underset{k_1}\to{\sum}\lambda_{kk_1}\, q_{k_1} \quad ,\\
  (quadratic) \quad q_k \mapsto q_k,  \qquad &p_k \mapsto p_k + 
                              \underset{k_1,k_2}\to{\sum}\lambda_{kk_1k_2}\, :q_{k_1}q_{k_2}: \quad ,\\
  (polynomial) \quad q_k \mapsto q_k,  \qquad &p_k \mapsto p_k + 
                              \underset{k_1,\ldots,k_n}\to{\sum}\lambda_{kk_1\cdots k_n}\, :q_{k_1}\cdots q_{k_n}: \quad ,
\endalign$$

\nind with all coefficients totally symmetric in the indices.\footnote{In point
of fact, the polynomial representations of the CCR constructed in \cite{30}
were constructed globally and not via their infinitesimal generators as is 
done in this paper.} The canonical 
transformations of general degree we construct in this paper are of the form

$$ \qquad q_k \mapsto q_k,  \qquad p_k \mapsto p_k + 
         \underset{m,k_1,\ldots,k_m}\to{\sum}
\lambda_{kk_1 \cdots k_m}\, :q_{k_1}\cdots q_{k_m}: \quad . \tag{1.1}
$$
 
     We emphasize that this generalization opens up a much larger class of
representations of the CCR --- representations, which involve an extremely 
singular perturbation of the original Fock representation. Aside from the
mathematical interest in this extension of the representation theory of an
important class of $C^*$-algebras associated with the CCR, there are 
advantages for the
quantum theory of bosonic systems with infinitely many degrees of freedom. 
As alluded to earlier, such representations allow rigorous treatment of 
interacting systems with extremely singular ``bare'' interaction. One may
therefore handle systems with new classes of dynamics (for a suggestive
treatment of the connection between representation and dynamics, see the 
discussion in \cite{1}). This will entail that the 
divergences which typically arise in perturbation theory can be avoided, since
the very need for a perturbation expansion from the ``wrong'' representation
is obviated. Furthermore, as briefly shown in the final section of \cite{22},
but which is well-known to theoretical physicists, the Hamilton operator,
which represents the total energy of the system, can often be simplified and 
in certain situations even be diagonalized by such canonical transformations. 
This leads to significant computational advantages. We postpone addressing 
these issues in detail to a later publication. \par
     In Chapter II, the general setting of this work will be more precisely
specified. In particular, we define and collect necessary facts about the 
$C^*$-algebras associated with the CCR and about their Fock representations. 
The new transformations of general degree will be constructed in 
Section 3.1, and those which lead to regular representations of the CCR will be
characterized in a number of different manners --- see Theorems 3.1.5 and 
3.1.6. We generalize a result in \cite{22} by showing that the general 
degree transformations yielding regular representations of the CCR can be 
brought into the transparent form given in equation (1.1), which we call
the standard form of such transformations. This fact proves to be useful in
the proof of some of the later results. \par
     Convenient sufficient conditions assuring the irreducibility of such 
representations are given in Section 3.2 --- see Theorem 3.2.2. However, we
show in Section 3.2 that there do exist such unrestricted order 
representations which are reducible. Chapter IV is given over to a proof that 
the class of representations constructed in Section 3.1 subsumes the 
previously studied classes discussed above. In particular, it is explicitly
shown that every (irreducible) quasifree or coherent representation of the CCR 
is unitarily equivalent to one of our unrestricted order representations
(Theorem 4.5), since it is already evident that the quadratic and polynomial
representations are included. In the process, we provide a new 
characterization of quasifree states on CCR-algebras (Proposition 4.4). \par
     In Chapter V we restrict our attention to the computationally more
manageable canonical transformations of finite degree and prove  
necessary and sufficient conditions for such representations to be 
quasi-equivalent to Fock (Theorem 5.6), coherent or quasifree (Theorem 5.10
and Corollary 5.11) representations. For our purposes, the primary interest of 
these results is in the violation of the identified necessary and sufficient 
conditions --- for then one has representations of bosonic systems of 
infinitely many degrees of freedom which describe {\it new} physics, 
{\it i.e.} which model bosonic systems that cannot be described by the earlier 
classes of representations. However, in this paper we do not 
try to study the new physical content of these representations. As an aside, 
the previously known conditions for unitary equivalence of quasifree states
follow as a special case of the results in Chapter V --- see Theorem 5.12.
The significantly less transparent conditions for unitary equivalence with the 
quadratic representations constructed in \cite{22} will not be given here, 
though they are known to us. (See the end of Chapter V for a brief 
discussion.)\par
     As this paper is an extension of \cite{22} and employs a number of the
results and arguments from that paper, we shall maintain the same notational
conventions, as detailed in the next chapter. Early versions of some of the 
results presented in this paper appeared in the {\it Diplomarbeit} \cite{11}.

\bigpagebreak

\heading II. Notation and General Setting \endheading

     We begin with an arbitrary real nondegenerate symplectic space
$(H,\sigma)$ with an associated regular Weyl system $(\Ks,W(f))$ consisting 
of a complex Hilbert space $\Ks$ and a mapping $W: H \rightarrow \Us(\Ks)$ from 
$H$ into the group $\Us(\Ks)$ of unitary operators on $\Ks$ which satisfies the
following axioms \cite{16}:     \par

$$W(f)W(g) = e^{-i\sigma(f,g)/2}W(f+g), \quad \forall f,g \in H, \tag{2.1}$$
$$W(f)^* = W(-f), \quad \forall f \in H, \tag{2.2}$$
\nind and 
$$\RR \ni t \mapsto W(tf) \in \Bs(\Ks) \quad 
\text{is weakly continuous for all} \quad f \in H . \tag{2.3}$$
\nind ($\Bs(\Ks)$ denotes the set of all bounded linear operators
$A:\Ks \rightarrow \Ks$.) Condition (2.3) entails that the map 
$\RR \ni t \mapsto W(tf) \in \Bs(\Ks)$ is actually strongly continuous, hence 
by Stone's Theorem one knows that for each $f \in H$ there exists a 
self-adjoint operator $\Phi(f)$ on $\Ks$ such that $W(tf) = e^{it\Phi(f)}$ for 
all $t \in \RR$ and by (2.1) the map $f \mapsto \Phi(f)$ is (real) linear. In 
fact, there exists a dense domain of vectors $D_W \subset \Ks$ which is a core 
of and left invariant by every $\Phi(f)$ \cite{23}\cite{13}; it is on 
this domain that the linearity just mentioned can be verified. On this domain 
one also verifies that the generators satisfy the CCR:     \par

$$\Phi(f)\Phi(g) - \Phi(g)\Phi(f) = i\sigma(f,g)\idty, \quad \forall f,g \in H. 
\tag{2.4}$$

\nind We therefore also call the Weyl system $(\Ks,W(f))$ and its associated 
generators as above a regular representation of the CCR over $(H,\sigma)$. 
In the physical literature such infinitesimal generators $\Phi(f)$ are
called field operators, and we shall use this language in the following. \par
     We shall denote by $\As(H,\sigma)$ the $C^*$-algebra on $\Ks$ generated 
by the operators $\{W(f) \mid f \in H\}$. As the notation indicates, the 
algebra $\As(H,\sigma)$ does not depend on the choice of representation of 
the Weyl operators $\{W(f) \mid f \in H\}$ (\cite{36} or Theorem 5.2.8 in 
\cite{6}). $\As(H,\sigma)$ is a simple $C^*$-algebra\footnote{The
fact that the CCR-algebra is simple can be seen as the correct generalization
of the Stone-von Neumann uniqueness theorem to the case of infinitely many
degrees of freedom. Indeed, since $\As(H,\sigma)$ is simple, all of its 
representations are isomorphic. When $H$ is finite-dimensional, this 
isomorphism is unitarily implementable, entailing the result in \cite{20}.}
and is nonseparable if $H$ is infinite-dimensional \cite{16}\cite{36}. There 
are, in fact, many 
different $C^*$-algebras one can associate with the CCR (see \cite{14} 
for a discussion of some of the alternatives), and the one we have chosen is 
minimal in the sense of set containment \cite{19}; but, for practical 
purposes the choice is immaterial, since one is generally interested in a 
von Neumann algebra which is `generated' by the $C^*$-algebra, and all the 
$C^*$-algebras discussed in \cite{14} (realized concretely on a given 
representation space) have the same weak closure. In the following it shall
be understood that one choice of $\sigma$ has been fixed, and we shall write
$\As(H)$ instead of $\As(H,\sigma)$. \par
     For any real linear map $J:H \rightarrow H$ satisfying $\sigma(Jf,Jg) =
\sigma(f,g)$, $-\sigma(Jf,f) > 0$ ($f \neq 0$), and $J^2 = -\idty$, one can 
introduce a complex structure on $H$ as follows \footnote{It should be 
mentioned that such a $J$ does not necessarily exist for arbitrary choice of 
symplectic space $(H,\sigma)$ \cite{28}.} \cite{14}: 
$(\alpha + i\beta)f \equiv \alpha f + \beta Jf$, for all $f \in H$, and
$\alpha,\beta \in \RR$. Moreover, 
$\langle f,g\rangle_{\Hs} \equiv -\sigma(Jf,g) + i\sigma(f,g)$ defines a 
scalar product on $H$ such that $(H,\langle\cdot,\cdot\rangle_{\Hs})$ is a 
complex preHilbert space \cite{14}, the completion of which we shall 
denote by $\Hs$. If, on the other hand, one begins with a complex Hilbert 
space $\Hs$ with scalar product $\langle\cdot,\cdot\rangle_{\Hs}$, then with 
$\sigma(f,g) = \Im m \langle f, g\rangle_{\Hs}$, $(\Hs,\sigma)$ is a real 
symplectic space of the sort with which we began, and with 
$\langle f,g\rangle \equiv \Re e \langle f, g\rangle_{\Hs}$, then 
$(\Hs,\langle\cdot,\cdot\rangle)$ is a real Hilbert space.  \par
     With a choice of a $\sigma$-admissible complex structure $J$ on 
$(H,\sigma)$, there is an important representation of $\As(H)$ called 
the Fock representation. This is given as the GNS-representation 
$(\Ks,\pi_J,\Omega)$ associated with the state \cite{18} $\omega_J$ 
determined on $\As(H)$ by

$$\omega_J(W(f)) \equiv e^{\sigma(Jf,f)/4}. \tag{2.5}$$

\nind Given such a representation, one can define the following `annihilation' 
and `creation' operators: 

$$a(f) \equiv \frac{1}{\sqrt{2}}(\Phi(f) + i\Phi(Jf)), \qquad  
      a^{\da}(f) \equiv \frac{1}{\sqrt{2}}(\Phi(f) - i\Phi(Jf)), \tag{2.6}$$

\nind where $\pi_J(W(tf)) = e^{it\Phi(f)}$. One has then $a(f)\Omega = 0$ for all
$f \in H$. \par
     For the purposes of this paper, we shall assume that a choice of 
$\sigma$-admissible complex structure $J$ has been made on $(H,\sigma)$ and
held fixed, so we 
have the complex one-particle space $\Hs$ and a corresponding Fock 
representation. Since as sets $H = \Hs$, we shall distinguish notationally the 
vector $f$ viewed as an element of the real Hilbert space $H$ from the same 
vector, denoted as $\tilde{f}$, viewed as an element of the complex Hilbert 
space $\Hs$. If $\{\tilde{e}_k \}_{k\in\IN}$ forms an orthonormal basis in 
$\Hs$, then the set $\{e_k, Je_k\}_{k\in\IN}$ forms a symplectic orthonormal 
system in $H$, in particular
$$\align
\sigma(e_k,e_l) = \sigma(Je_k,Je_l) = 0 \quad &, \quad
   \sigma(e_k,Je_l) = \delta_{kl} \quad , \\
\langle e_k, e_l \rangle = \langle Je_k, Je_l \rangle =
  \delta_{kl} \quad &, \quad \langle e_k, Je_l \rangle = 0 \quad .
\endalign$$

\nind In this paper, whenever a choice of symplectic orthonormal basis 
$\{e_k, Je_k\}_{k\in\IN}$ has been made, the Hilbert subspace of $H$
generated by $\{e_k\}_{k\in\IN}$ will be denoted by $V$. \par
     With the choice of complex structure made as above, the corresponding 
Fock state $\omega_J: \As(H) \rightarrow \CC$ now satisfies

$$\omega_J(W(f)) = e^{-\Vert f \Vert^2 /4},$$

\nind and the associated GNS-space may be represented by the symmetric
Fock space $\Fs_+(\Hs)$. We recall that the Fock space 
${\Fs}({\Hs}) = \bigoplus_{n=0}^{\infty} {\Hs}^n$ (${\Hs}^0 = \CC$, 
${\Hs}^n$ is the $n$-fold tensor product of $\Hs$ with itself), and that 
$\Fs_+(\Hs)$ is the totally symmetric subspace 
$\bigoplus_{n=0}^{\infty}P_+ {\Hs}^n $ of ${\Fs}({\Hs})$, where $P_+$ is the 
projection

$$P_+ (\tilde{f}_1 \otimes \tilde{f}_2 \otimes \cdots \otimes \tilde{f}_n ) =
\frac{1}{n!} \sum_{\pi} \tilde{f}_{\pi(1)} \otimes \tilde{f}_{\pi(2)} \otimes
\cdots \otimes \tilde{f}_{\pi(n)}$$

\nind ($\tilde{f}_i \in {\Hs}$, $\pi \in S_n$, the group of permutations on 
the set $\{1, 2,\ldots, n \}$). The projection operator 
$P_n : \FH \rightarrow \Hs^n$ projects onto the n-particle subspace. 
The vector $\Omega \equiv (1, 0, 0, \ldots) \in \Fs_+(\Hs)$ is 
the Fock vacuum. $\Fo$ is the finite-particle subspace, i.e. the linear span 
of the ranges of $\{P_n \}_{n \in \IN_0}$. The elements of $\Fs_0$ will be 
called finite-particle vectors. As is well-known, the GNS-representation for 
the Fock state may be identified with this Fock space representation. \par
     With $a(\tilde{f})$ the usual annihilation operator in $\Fs_+(\Hs)$ for 
$\tilde{f} \in \Hs$ and the adjoint operator $a^*(\tilde{f})$ (an
extension of) the
corresponding creation operator, then the linear self-adjoint operator
$\Phi_S(\tilde{f}) \equiv \frac{1}{\sqrt{2}} \, (\overline{a(\tilde{f}) +
a^*(\tilde{f})})$ (the bar denotes the closure of the operator) is called 
the Segal field operator in $\Fs_+(\Hs)$. If we also view $\tilde{f} \in \Hs$ 
as an element of $H$, then we have $\Phi_S(\tilde{f}) = \Phi(f)$, after 
identifying $\As(H)$ with $\pi_J(\As(H))$ (since they are isomorphic). $\Fo$ 
is a core for $\Phi(f)$, and for $\varphi \in \Fs_0$, $\Phi(f)$ satisfies the 
bound $\Vert \Phi(f) \varphi \Vert \leq \sqrt{2} \sqrt{\nphi+1} \Vert f \Vert 
\Vert\varphi\Vert$, where $\nphi $ equals the smallest $n \in \IN$ such that 
$P_N\varphi = 0$ for all $N > n$. If $\{ f_n \}_{n=1}^\infty $ converges in 
$H$ to $f$, then for every $\varphi\in \Fo$ the sequence 
$\{\Phi(f_n) \varphi\}_{n=1}^\infty $ converges in $\Fs_+(\Hs)$ to 
$\Phi(f) \varphi$. \par
To make a notational connection to the discussion in Chapter I and in keeping 
with common harmonic oscillator conventions, if $\{e_k,Je_k\}$ is a symplectic
orthonormal basis in $H$, then the `position operator' and the 
`momentum operator' corresponding to the $k$-th degree of freedom are given 
by

$$q_k \equiv \Phi_S(\tilde{e}_k) = \Phi(e_k) \quad , \quad
  p_k \equiv \Phi_S(i\tilde{e}_k) = \Phi(J e_k) \quad .$$

     To elucidate what is meant in the following by linear canonical 
transformations, we recall two well-studied classes of representations of the
CCR --- the coherent and the quasifree representations. If $(\Ks,W(f))$ is a 
representation of $\As(H)$ and $l : H \rightarrow \RR$ is a linear map, then

$$\pi_l(W(f)) \equiv \hat{W}(f) \equiv e^{i\,l(f)}W(f) \quad , \quad f \in H \quad , $$

\nind determines a representation of $\As(H)$ generally called a coherent 
representation if $(\Ks,W(f))$ is a Fock representation (see, {\it e.g.}
\cite{31}). This leads to the relationship  
$\hat{\Phi}(f) = \Phi(f) + l(f)\idty,\, f \in H$, between the generators 
of the representations. One may equivalently start with a Fock state 
$\omega_J$ on $\As(H)$ with associated representation $(\Ks, \pi_J )$ and 
define a coherent state $\omega_l$ by

$$\omega_l(W(f)) \equiv \omega_J(W(f))e^{i\, l(f)}\quad , \quad f \in H \quad .$$

\nind Of course, the GNS representation of $\As(H)$ corresponding to
$\omega_l$ is given on $\Ks$ by 
$$\pi_l(W(f)) \equiv e^{i\, l(f)}\pi_J(W(f)) \quad , \quad f \in H \quad . $$ 

\nind Quasifree (or symplectic) representations can be obtained from a given 
Fock representation of the CCR-algebra $\As(H)$ by a Bogoliubov transformation 
$$\align 
\Phi_T(f) &\equiv \Phi(Tf) = \Phi(f) + \Phi((T-\idty)f) \quad , \\
(\text{or equivalently} \quad \pi_T(W(f)) &\equiv \pi_J(W(Tf)) \quad , \endalign $$ 
\nind using a symplectic operator $T$, i.e.
one leaving the symplectic form invariant: $\sigma(Tf,Tg) = \sigma(f,g)$,
for all $f,g \in H$. The canonical transformations associated with the 
coherent and quasifree representations constitute the inhomogeneous linear 
group of canonical transformations, studied by Shale \cite{33}
and Berezin \cite{4}, among many others. \par
     A crucial technical tool in this paper is the use of $Q$-space techniques.
Let $\{e_k,Je_k\}$ be a
symplectic orthonormal basis. Moreover, let $x = (x_{1}, x_{2}, \ldots )$ be a 
point in $Q \equiv \times_{k=1}^{\infty}\ \RR $, and $\Sigma$ be the 
$\sigma$-algebra generated by the cylinder sets of $Q$ with Lebesgue 
measurable base. Then $\mu = \otimes_{k=1}^{\infty} \ \mu_{k} $,  where each 
$\mu_{k}$ is the Gaussian measure 
$d\mu_{k} = \pi^{-\frac{1}{2}} e^{-x_{k}^{2}}\, dx_{k} $, is a probability 
measure on $( Q, \Sigma ) $. It is well-known that there exists a unitary map 
$S$ of $\Fs_+(\Hs)$ onto $L^{2}(Q,d\mu)$ such that \cite{32}\cite{35}\cite{25} 
$$\align
S \Omega_{0}  =  1  \quad &\text{and} \quad
  S\, P_{+}\,(\tilde{e}_{k_{1}}\otimes \tilde{e}_{k_{2}}\otimes \cdots 
\otimes \tilde{e}_{k_{r}}) = 
  (r!)^{-\frac{1}{2}} (\sqrt{2})^{r} : x_{k_{1}} x_{k_{2}} \cdots 
  x_{k_{r}} : \ ,  \\
S \Phi(e_{k}) S^{-1} &\equiv q_{k} = x_k \idty = \ \text{(multiplication by)} \ \ x_{k} ,
\ \  \text{and} \\
S \Phi(J e_{k}) S^{-1} &\equiv p_{k} 
   = \ \frac{1}{i} \frac{\partial}{\partial x_{k}} + i x_{k} \idty \ ,
\endalign $$
 
\nind where the operator equations are understood to hold on the dense set
$S\Fs_0$. We shall drop the symbol $S$, when the identification is 
clear. \par
     Let $\underline{k} = \{k_{1},k_{2},\ldots , k_{r} \}$ be a multiple index 
in $\IN ^r = \times_{j=1}^{r}\IN$,
$\{\lambda_{\underline{k}}\}_{\underline{k} \in \IN ^r}$
a sequence of real numbers, totally symmetric in $\underline{k}$ , 
$\sum_{\underline{k}}\lambda_{\underline{k}}^{2} < \infty$,
and \newline $I_{n} = \{ \underline{k} \mid 
\max \{k_{1},k_{2},\ldots,k_{r}\} \leq n \}$. Then 

$$ f_{n}^{(r)} = \sum_{\underline{k}\in I_{n}} \lambda_{\underline{k}} \ 
e_{k_{1}}\otimes e_{k_{2}}\otimes \cdots \otimes e_{k_{r}} \ \in 
H^r \, ,$$

\nind with $\| f_{n}^{(r)} \|^{2} = 
\sum_{\underline{k}\in I_{n}} \lambda_{\underline{k}}^{2}$. Since with 
$m < n$, 
we have $\| f_{n}^{(r)} - f_{m}^{(r)} \|^{2} = \sum_{\underline{k} \in I_{n}
 \backslash I_{m}} \lambda_{\underline{k}}^{2} $
and $ \sum_{\underline{k}\in I_{n}} \lambda_{\underline{k}}^{2}$ 
converges with $n \rightarrow \infty $, it follows that

$$f_{n}^{(r)} \ \longrightarrow \ f^{(r)} = 
\sum_{\underline{k}} \lambda_{\underline{k}} \ 
e_{k_{1}}\otimes e_{k_{2}}\otimes \cdots \otimes e_{k_{r}} \ \ \text{in} \  
H^r \ . $$

   Consider the sequence of operators

$$A(f_{n}^{(r)}) =  \sum_{\underline{k}\in I_{n}} \lambda_{\underline{k}}\,
: \Phi(e_{k_{1}}) \Phi(e_{k_{2}}) \cdots \Phi(e_{k_{r}}) :\ , $$

\nind the $Q$--space realization of which is given by
$$A_{n} = A(f_{n}^{(r)}) =\sum_{\underline{k}\in I_{n}} 
\lambda_{\underline{k}}\,: x_{k_{1}} x_{k_{2}} \cdots x_{k_{r}} : \idty \quad .$$
Note that since it is a polynomial, $A_{n} = A_{n}(x)$ is in $L^{2}(Q,d\mu)$ 
with
$$\align
\| A_{n} \|^{2} &=  2^{-r} \ r! \
\sum_{\underline{k}\in I_{n}} \lambda_{\underline{k}}^{2} =
2^{-r} \ r! \ \| f_{n}^{(r)} \|^{2} \ \ , 
 \quad \\
\text{and} \qquad \| A_{n} - A_{m} \|^{2} &= 2^{-r} \ r! \ \
\| f_{n}^{(r)} - f_{m}^{(r)} \|^{2}    
\endalign $$

\nind ( cf. the proof of Lemma I.18 of \cite{35} ).
Therefore, $A_{n}(x)$ converges in $L^{2}(Q,d\mu)$, and we shall call the 
a.e.-defined limit $A(x) = \sum_{\underline{k}} \lambda_{\underline{k}}\,
: x_{k_{1}} x_{k_{2}} \cdots x_{k_{r}} : $, which up to a factor of 
$\sqrt{r!} \, (\sqrt{2})^{-r}$ corresponds to $f^{(r)}$ . \par
    The advantage of the $Q$--space formulation is that all functions of the
elements of $ \{ \Phi (e_{k}) \mid k \in \IN \}$ become multiplication 
operators on $L^{2}(Q,d\mu)$. $A_{n}(x)$ and $A(x)$ are measurable, 
real-valued functions on $Q$ which are finite almost everywhere with respect 
to $\mu$. So with $D(A) \equiv \{ \varphi \mid A\varphi \in L^{2}(Q,d\mu) \} $ 
(similarly for $D(A_n)$), $ (A_{n} \varphi)(x) = A_{n}(x) \varphi (x) $ and
$(A \varphi)(x) = A(x) \varphi (x) $ are self-adjoint operators ( cf. 
\cite{24}, VIII.3 \ Proposition 1 ). Thus, for every 
$f^{(r)} \in P_{+} \, V^r$, $A(f^{(r)})$ 
represents a well-defined self-adjoint multiplication operator. 

\bigpagebreak

\heading III. A General Class of Representations of the CCR \endheading

     In Section 3.1 we shall use $Q$-space techniques to construct our 
general class of representations of the CCR, as motivated above. These 
techniques were already employed to establish some of the results proven in 
\cite{22} about quadratic representations of the CCR. Our representations
of general degree of the CCR will be defined in terms of linear maps $\Lambda$
from the one-particle space $H$ into $Q$-space itself. We shall give a number
of characterizations of those $\Lambda$ which yield representations of
the CCR, {\it i.e.} which determine canonical transformations of the field
operators. It will be shown in Section 3.2 that when such $\Lambda$ are
bounded, then the resulting representation is irreducible. We shall also
explain how unbounded $\Lambda$ can lead to reducible representations.

\proclaim{3.1 Canonical Transformations of Arbitrary Degree}
\endproclaim 

We shall use the basic facts that the set
$\{ 1 \} \cup \{\, :x_{k_1} \ldots x_{k_m}: \,\mid k_1,\ldots,k_m,m\in\IN\,\}$ 
is an orthogonal basis in $L^2(Q,d\mu)$ and that

$$\Vert \underset{k_1,\ldots,k_m}\to{\sum} \lambda_{k_1 \ldots k_m}
:x_{k_1} \ldots x_{k_m}: \Vert_2^2 = \underset{k_1,\ldots,k_m}\to{\sum}
\lambda^2_{k_1 \ldots k_m}\frac{m!}{2^m} \quad , \tag{3.1.1} $$

\nind for arbitrary $\lambda_{k_1 \ldots k_m} \in \RR$ symmetric in the
indices $k_1 \ldots k_m$ (see Section 4.3 in \cite{22}). Thus, standard
arguments entail the following lemma. 

\proclaim{Lemma 3.1.1} Let 
$F = \underset{k_1,\ldots,k_m}\to{\sum} c_{k_1 \ldots k_m}
:x_{k_1} \ldots x_{k_m}: \,\in \L2Q$ with complex numbers $c_{k_1 \ldots k_m}$ 
symmetric in the indices $k_1 \ldots k_m$. Then 

$$\langle F, :x_{k_1} \ldots x_{k_m}: \rangle = 
c_{k_1 \ldots k_m}\frac{m!}{2^m} \quad , $$

\nind and $F$ is real-valued (a.e. $\mu$) if and only if 
$c_{k_1 \ldots k_m} \in \RR$ for all $k_1,\ldots,k_m \in \IN$. 
\endproclaim

     We also state a well-known fact about the $Q$-space representation
of Fock space. Recall that for $f \in V$, $x(f)\idty = \Phi(f)$, given our
stated convention of dropping the unitary $S$. The number operator $N$
on Fock space is a self-adjoint operator satisfying 
$N P_n \varphi = n P_n \varphi$, for all $\varphi \in \Fs(\Hs)$ and commutes
with $P_+$.

\proclaim{Lemma 3.1.2} The set $\Gs$ equal to the linear span of 
$\{ e^{ix(f)} \mid f \in V \}$ is a core for the number operator $N$ and for 
$\Phi(g)$, given any $g \in H$.
\endproclaim 

     In the next proposition we relate the operators $\Phi(f)$ and
$\Phi(f) + F\idty$, for any $F \in \L2Q$, via a unitary transformation.

\proclaim{Proposition 3.1.3} If $F \in \L2Q$ is real-valued ($\mu$ a.e.), then
the operator $\Phi(f) + F\idty$, is essentially self-adjoint on
$L^{\infty}(Q,d\mu)$, when $f \in V$, and on $e^{-iG}\Gs$, when $f \not\in V$ 
($G$ will be defined shortly).
Furthermore, if $f \not\in V$, then the closure of the corresponding operator
is given in terms of the self-adjoint $\Phi(f)$ by
$$\overline{\Phi(f) + F\idty} = e^{-iG}\Phi(f)e^{iG} \quad , \tag{3.1.2}$$

\nind for any choice of ($\mu$ a.e.) real-valued $G \in \L2Q$ such that
$\partial G / \partial x(f_2) = F$, where $f_2$ is determined uniquely by
$f$ by the decomposition $f = f_1 + Jf_2$, $f_1,f_2 \in V$.
\endproclaim

\demo{Proof} By Lemma 4.3.1 in \cite{22}, $L^{\infty}(Q,d\mu)$ is a core
for the corresponding multiplication operator for every $F \in \L2Q$. If
$f \in V$, then the operator $\Phi(f) + F\idty$ is symmetric,
$D(\Phi(f) + F\idty) = D(\Phi(f)) \cap D(F\idty)$, and, on 
$L^{\infty}(Q,d\mu) \subset \L2Q$, it is
equal to the operator  corresponding to multiplication by the $L^2$-function
$x(f) + F$, {\it i.e.} $\Phi(f) + F\idty$ is essentially self-adjoint on 
$L^{\infty}(Q,d\mu)$. \par
     One may assume that $f \not \in V$ and therefore $f = Je_1 + v$ for a
suitable $v \in V$ (after choosing the basis $\{ e_k \mid k \in \IN\}$
appropriately). There exist suitable $G_n$, $n \in \IN\cup\{ 0 \}$, in the
subspace of $\L2Q$, which includes the constant functions and the set
$\{ :x_{k_1} \ldots x_{k_m}: \mid k_1,\ldots,k_m \geq 2 \}$, such that

$$F = \underset{n = 0}\to{\overset{\infty}\to{\sum}} :x_1^n:G_n \quad . $$

\nind Set $G = \sum_{n=0}^{\infty} \frac{1}{n+1} :x_1^{n+1}: G_n$. The series
converges with respect to the $L^2$-norm, {\it i.e.} $G \in \L2Q$, since
$$\align
\Vert G \Vert_2^2 \quad &= \quad \underset{n}\to{\sum} \frac{1}{(n+1)^2} 
\Vert :x_1^{n+1}:\Vert_2^2 \, \Vert G_n \Vert_2^2 \\
&\overset{(3.1.1)}\to{=} \underset{n}\to{\sum} \frac{1}{(n+1)^2} 2^{-n-1}(n+1)!
\Vert G_n \Vert_2^2 \\
&\leq \quad\underset{n}\to{\sum} \, 2^{-n}n! \Vert G_n \Vert_2^2 \\
&= \quad\underset{n}\to{\sum}\, \Vert :x_1^{n}:\Vert_2^2 \, \Vert G_n \Vert_2^2 \\
&= \quad\Vert F \Vert_2^2 < \infty \quad ,  
\endalign $$

\nind and $G,G_n$, $n \in \IN$, are ($\mu$-a.e.) real-valued, by Lemma 3.1.1.
\par
     Note that $\partial : x_1^k :/\partial x_1 = k:x_1^{k-1}:$ is a 
polynomial of degree $k-1$ with leading term $kx_1^{k-1}$. It is evident that
$\Gs \subset \LiQ$, but one has, furthermore, $e^{-iG}\Gs \subset D(\Phi(f))$
and $\Phi(f)e^{-iG}\varphi = e^{-iG}(\Phi(f)-F\idty)\varphi$, for all 
$\varphi \in \Gs$, since

$$e^{-i\sum_{n=0}^m \frac{1}{n+1} :x_1^{n+1}:G_n} \rightarrow
e^{-iG}$$

\nind strongly as $m \rightarrow \infty$ (see the proof of Lemma 4.3.1 in
\cite{22}) and
$$\align
\Phi(f) e^{-i\sum_{n=0}^m \frac{1}{n+1} :x_1^{n+1}:G_n} \varphi \quad &= \quad
(\frac{\partial}{i\partial x_1} + ix_1 + x(v))
e^{-i\sum_{n=0}^m \frac{1}{n+1} :x_1^{n+1}:G_n} \varphi \\
&= \quad e^{-i\sum_{n=0}^m \frac{1}{n+1} :x_1^{n+1}:G_n}(\Phi(f) - 
\underset{n=0}\to{\overset{m}\to{\sum}}\, :x_1^n :G_n)\varphi \\
&\rightarrow \quad e^{-iG}(\Phi(f)-F\idty)\varphi 
\endalign $$

\nind with respect to the $L^2$-norm. Therefore, 
$$\align
e^{iG}(\Phi(f) + F\idty)e^{-iG}\varphi \quad &= \quad e^{iG}\Phi(f)e^{-iG}\varphi +
F\varphi \\
&= \quad (\Phi(f) - F\idty)\varphi + F\varphi \\
&= \quad \Phi(f)\varphi \quad ,
\endalign $$

\nind for any $\varphi \in \Gs$. Since $\Gs$ is a core for the self-adjoint
operator $\Phi(f)$, the symmetric operator $\Phi(f) + F\idty$ is essentially
self-adjoint on $e^{-iG}\Gs$ and (3.1.2) holds. \par
     To see the truth of the final assertion of the proposition, note that
if $G_0 \in \L2Q$ satisfies $\frac{\partial}{\partial x_1}G_0 = F$, then
$G - G_0$ is an element of the subspace generated by
$\{ :x_{k_1}\cdots x_{k_m}: \mid k_1,\ldots,k_m \geq 2 \} \subset \L2Q$,
{\it i.e.} $\Phi(f)$ commutes with $e^{i(G - G_0)}$.
\hfill\qed\enddemo

     We therefore see that for any real-linear, densely defined
$\Lambda : H \rightarrow \L2Q$ we obtain self-adjoint transforms of the
field operators $\Phi(f)$:

$$\Phi_{\Lambda}(f) \equiv \overline{\Phi(f)+\Lambda f \idty} \quad , \quad
f \in D(\Lambda) \quad , $$

\nind where the closure is understood to be taken on 
$D(\Phi(f)) \cap D(\Lambda f \idty)$. It is, of course,
not true in general that the operators 
$\{\Phi_{\Lambda}(f) \mid f \in D(\Lambda)\}$ form a representation of the 
CCR. We shall concentrate upon those which do.

\proclaim{Definition 3.1.4} Let $\Ls$ be the set of all real-linear, densely
defined maps from $H$ to $\L2Q$, and let $\Ls_{CCR} \subset \Ls$ be the 
subset of $\Ls$ consisting of elements $\Lambda$ such that

$$\pi_{\Lambda}(W(f)) \equiv e^{i\Phi_{\Lambda}(f)} \quad , \quad f \in D(\Lambda) \quad ,$$

\nind defines a regular representation $(\pi_{\Lambda},\L2Q)$ of the
CCR-algebra $\As(D(\Lambda))$.
\endproclaim

\nind Note that, by Corollary 4.1.2 in \cite{22}, this definition generalizes
the one made in \cite{22}. \par
     Hence, each $\Lambda \in \Ls_{CCR}$ induces a canonical transformation
on the quantum fields $\Phi(f) \mapsto \Phi_{\Lambda}(f)$, which itself is
exponentiable to yield a regular representation of the algebra 
$\As(D(\Lambda))$. We now wish to characterize the members of this set 
$\Ls_{CCR}$ and shall do so in more than one way. Let 
$P_n : \L2Q \rightarrow \L2Q$, $n \in \IN$, be the orthogonal projection onto the 
subspace of $n$-particle vectors and $P : H \rightarrow H$ be the orthogonal 
projection onto the subspace $JV$. 

\proclaim{Theorem 3.1.5} Let $\Lambda \in \Ls$. Then $\Lambda \in \Ls_{CCR}$ 
if and only if $\Lambda h$ is a ($\mu$ a.e.) real-valued function, for each
$h \in D(\Lambda)$, and 
$$\langle \Lambda f,a^*(Pg)\psi\rangle = \langle \Lambda g ,a^*(Pf)\psi\rangle
\quad , \tag{3.1.3}$$
\nind for arbitrary $f,g \in D(\Lambda)$ and 
$\psi \in D(a^*(Pf))\cap D(a^*(Pg))$. The assertion still holds if (3.1.3) is
replaced by
$$a(Pf)P_n\Lambda g = a(Pg)P_n \Lambda f \quad , \tag{3.1.4}$$
\nind for all $f,g \in D(\Lambda)$ and $n \in \IN$.
\endproclaim

\demo{Proof} Assume that $\Lambda \in \Ls_{CCR}$. Then because the field
operator $\Phi_{\Lambda}(h)$ must be self-adjoint, the function $\Lambda h$
must be real-valued ($\mu$ a.e.). Equations (3.1.3) and (3.1.4) are trivial
if $f,g \in V$, so that $Pf = 0 = Pg$. Hence, one may assume that
$f = Je_1 + v_1$ and $g = c_1Je_1 + c_2Je_2 + v_2$ for suitable 
$c_1,c_2 \in \RR$ and $v_1,v_2 \in V$, after choosing the basis
$\{ e_k,Je_k \mid k \in \IN \}$ appropriately. Differentiating the equation

$$\langle e^{it\Phi_{\Lambda}(f)}\varphi,e^{is\Phi_{\Lambda}(g)}\psi\rangle
= e^{its\sigma(f,g)}\langle e^{is\Phi_{\Lambda}(g)}\varphi,e^{it\Phi_{\Lambda}(f)}\psi\rangle $$

\nind with respect to $s$ and $t$ and evaluating at $t = 0 = s$, one can 
conclude, using Theorem VIII.7 in \cite{24} that

$$\langle \Phi_{\Lambda}(f)\varphi,\Phi_{\Lambda}(g)\psi\rangle =
\langle \Phi_{\Lambda}(g)\varphi,\Phi_{\Lambda}(f)\psi\rangle + i\sigma(f,g)
\langle\varphi,\psi\rangle \quad ,$$

\nind for any 
$\varphi, \psi \in \Gs \subset D(\Phi_{\Lambda}(f))\cap D(\Phi_{\Lambda}(g))$.
Since $\Lambda f$ and $\Lambda g$ are $\mu$ a.e. real-valued, one has
$$\align
0 &= \langle\Phi_{\Lambda}(f)\Omega,\Phi_{\Lambda}(g)\psi\rangle -
  \langle\Phi_{\Lambda}(g)\Omega,\Phi_{\Lambda}(f)\psi\rangle -
i\sigma(f,g)\langle\Omega,\psi\rangle \\
&= \langle\Phi(f)\Omega,\Phi(g)\psi\rangle + 
  \langle\Lambda f\Omega,\Phi(g)\psi\rangle +
  \langle\Phi(f)\Omega,\Lambda g\psi\rangle + 
  \langle\Lambda f\Omega,\Lambda g\psi\rangle \\
&\quad - \langle\Phi(g)\Omega,\Phi(f)\psi\rangle -
  \langle\Lambda g\Omega,\Phi(f)\psi\rangle -
  \langle\Phi(g)\Omega,\Lambda f\psi\rangle - 
  \langle\Lambda g\Omega,\Lambda f\psi\rangle \\
&\quad - i\sigma(f,g)\langle\Omega,\psi\rangle \\
&= \langle\Phi(f)\Omega,\Phi(g)\psi\rangle -
\langle\Phi(g)\Omega,\Phi(f)\psi\rangle -
i\sigma(f,g)\langle\Omega,\psi\rangle \\
&\quad + \langle\Lambda f\Omega,\Phi(g)\psi\rangle +
  \langle\Phi(f)\Omega,\Lambda g\psi\rangle -
  \langle\Phi(g)\Omega,\Lambda f\psi\rangle - 
  \langle\Lambda g\Omega,\Phi(f)\psi\rangle \\ 
&\quad + \langle\Lambda f\Omega,\Lambda g\psi\rangle -
  \langle\Lambda g\Omega,\Lambda f\psi\rangle \\
&= \langle\Lambda f\Omega,\Phi(g)\psi\rangle +
  \langle\Phi(f)\Omega,\Lambda g\psi\rangle -
  \langle\Phi(g)\Omega,\Lambda f\psi\rangle - 
  \langle\Lambda g\Omega,\Phi(f)\psi\rangle \\ 
&= \langle\Lambda f,(\Phi(g)-\overline{\Phi(g)\Omega}\idty)\psi\rangle -
   \langle\Lambda g,(\Phi(f)-\overline{\Phi(f)\Omega}\idty)\psi\rangle \quad .
\endalign $$
\nind By using
$$\align
(\Phi(f)-\overline{\Phi(f)\Omega}\idty)\psi &= 
(\Phi(Je_1)-\overline{\Phi(Je_1)\Omega}\idty)\psi = (\Phi(Je_1)+ix_1)\psi \\
 &= (\Phi(Je_1)+i\Phi(e_1))\psi = \sqrt{2}a^*(Je_1)\psi \\
 &= \sqrt{2}a^*(Pf)\psi 
\endalign $$
\nind and the similar equality 
$(\Phi(g)-\overline{\Phi(g)\Omega}\idty)\psi = \sqrt{2}a^*(Pg)\psi$, one can 
conclude that (3.1.3) is fulfilled for $\psi \in \Gs$. \par
     The next step is to prove (3.1.3) for polynomials $\psi \in \L2Q$. Since
$\Gs$ is a core for $N$ (Lemma 3.1.2), for arbitrary $k_1,\ldots,k_m \in \IN$
there exists a sequence $\{\psi_n \}\subset \Gs$ which converges in
$\L2Q$ to $:x_{k_1} \cdots x_{k_m}:$ such that also the sequence 
$\{ N\psi_n\}$ converges in $\L2Q$ to $N:x_{k_1} \cdots x_{k_m}:$. For 
arbitrary $h \in H$ one has
$$\align
\Vert a^*(h)&(\psi_n - :x_{k_1} \cdots x_{k_m}:)\Vert^2 \leq \\ 
&\Vert h \Vert^2 \langle(\psi_n - :x_{k_1} \cdots x_{k_m}:),
(N+1)(\psi_n - :x_{k_1} \cdots x_{k_m}:)\rangle \rightarrow 0 \quad ,
\endalign $$
\nind as $n \rightarrow \infty$, so that (3.1.3) holds for
$\psi = :x_{k_1} \cdots x_{k_m}:$. It follows that
$$\align
\langle \Lambda f,a^*(Pg)\psi\rangle &=
  \underset{n=0}\to{\overset{\infty}\to{\sum}} \langle P_n \Lambda f,P_n a^*(Pg)\psi\rangle 
= \underset{n=1}\to{\overset{\infty}\to{\sum}} \langle P_n \Lambda f,a^*(Pg)P_{n-1}\psi\rangle \\
&= \underset{n=1}\to{\overset{\infty}\to{\sum}} \langle \Lambda f,a^*(Pg)P_{n-1}\psi\rangle 
= \underset{n=1}\to{\overset{\infty}\to{\sum}} \langle \Lambda g,a^*(Pf)P_{n-1}\psi\rangle \\
&= \langle \Lambda g,a^*(Pf)\psi\rangle \quad , \tag{3.1.5}
\endalign $$
\nind for $\psi \in D(a^*(Pf)) \cap D(a^*(Pg))$. The chain of equalities
(3.1.5) establishes the asserted equivalence of (3.1.3) and (3.1.4). \par
     Assume now that (3.1.3) holds for arbitrary $f,g \in D(\Lambda)$ and that
$\Lambda h$ is $\mu$ a.e. real-valued, for each $h \in D(\Lambda)$. The
assertion in this direction will be established in part by appealing to 
another characterization of $\Lambda \in \Ls_{CCR}$ appearing in Theorem 3.1.6,
which will be presented subsequently. In particular, here it will be shown 
that there exists an extension $\Lambda'$ of $\Lambda$ containing $V$ in its 
domain of definition and satisfying the hypothesis of Theorem 3.1.6. It will 
then follow from Theorem 3.1.6 that both $\Lambda'$ and $\Lambda$ are contained
in $\Ls_{CCR}$. \par
     Let $g \in V \cap D(\Lambda)$ be arbitrary but fixed. The set
$\{ Pf \mid f \in D(\Lambda) \}$ is dense in $JV$, so the set
$\{ a^*(Pf)\psi \mid \psi \in D(a^*(Pf))\cap D(a^*(Pg)),f \in D(\Lambda)\}$
is dense in the orthogonal complement of the set $\{\Omega\}$. Therefore,
$\Lambda g$ must be a multiple of $\Omega$, {\it i.e.} a constant function,
whenever $Pg = 0$, {\it i.e.} whenever $g \in V$ (use the hypothesis (3.1.3)).
Thus, the restriction of $\Lambda$ to $V \cap D(\Lambda)$ determines a
linear form $\ell : V \cap D(\Lambda) \rightarrow \RR$. But $\ell$ has a
linear extension $\ell'$ to $V$, so $\Lambda$ has a linear extension
$\Lambda'$ to $D(\Lambda) + V$ such that $\Lambda' \mid V = \ell'$ and
such that (3.1.3) holds with $\Lambda$ replaced by $\Lambda'$ (both sides
of (3.1.3) under this replacement are equal to zero for the additional vectors
$g \in V$ and constant $\Lambda g$). It shall be established that this
extension $\Lambda'$ fulfills the hypothesis of Theorem 3.1.6. \par
     For arbitrary $f_1,\ldots,f_m \in JD(\Lambda')\cap V$, one has

$$a^*(f_2):x(f_3)\cdots x(f_m): = \sqrt{2}:x(f_2)x(f_3)\cdots x(f_m): \quad , $$

\nind so that
$$\align
\langle\Lambda' Jf_1, :x(f_2)\cdots x(f_m):\rangle &= \frac{1}{\sqrt{2}}
\langle\Lambda' Jf_1, a^*(f_2):x(f_3)\cdots x(f_m): \rangle \\
&= -\frac{i}{\sqrt{2}}\langle\Lambda' Jf_1, a^*(Jf_2):x(f_3)\cdots x(f_m): \rangle \\
&\overset{(3.1.3)}\to{=} -\frac{i}{\sqrt{2}}\langle\Lambda' Jf_2, a^*(Jf_1):x(f_3)\cdots x(f_m): \rangle \\
&= \langle\Lambda' Jf_2, :x(f_1)x(f_3)\cdots x(f_m):\rangle \quad ,
\endalign $$
\nind and the conditions (i)-(iii) of Theorem 3.1.6 are satisfied for
$\Lambda'$.
\hfill\qed\enddemo

     We now give the announced second characterization of 
$\Lambda \in \Ls_{CCR}$.

\proclaim{Theorem 3.1.6} If $\Lambda \in \Ls_{CCR}$, then there exists an
extension $\Lambda' \in \Ls_{CCR}$ of $\Lambda$ with $V \subset D(\Lambda')$.
Moreover, if $V \subset D(\Lambda)$, then $\Lambda \in \Ls_{CCR}$ is
equivalent to the following three conditions: \par
   (i) The functions $\Lambda h$ are $\mu$ a.e. real-valued, for all
$h \in D(\Lambda)$. \par
   (ii) For arbitrary $f_1,\ldots,f_m \in JD(\Lambda)\cap V$,
$$\langle\Lambda Jf_1, :x(f_2)x(f_3)\cdots x(f_m):\rangle =
\langle\Lambda Jf_2, :x(f_1)x(f_3)\cdots x(f_m):\rangle \quad . \tag{3.1.6}
$$

   (iii) $\Lambda f$ is a (real) constant function for all $f \in V$.
\endproclaim

\demo{Proof} The proof of Theorem 3.1.5 shows that $\Lambda \in \Ls_{CCR}$
implies that there exists an extension $\Lambda'$ of $\Lambda$ which fulfills
conditions (i)-(iii) and $V \subset D(\Lambda')$. It therefore remains only
to prove that $V \subset D(\Lambda)$ and (i)-(iii) imply 
$\Lambda \in \Ls_{CCR}$. By applying a suitable coherent transformation,
it may be assumed that $\langle\Lambda h,\Omega\rangle = 0$, for all
$h \in D(\Lambda)$. By choosing a suitable basis $\{ e_k \mid k \in \IN\}$
of $V$, it may also be assumed that 
$\{ Je_k \mid k \in \IN \} \subset D(\Lambda)$. \par
     Now let $f,g \in D(\Lambda)$ be arbitrary but fixed. One can choose the 
basis of $V$ such that $f = c_1 Je_1 + v_1$, $g = c_2 Je_1 + c_3 Je_2 + v_2$,
with $c_1,c_2,c_3 \in \RR$ and $v_1,v_2 \in V$. There are suitable 
$\lambda_{k_1 \ldots k_m} \in \RR$ ($m,k_1,\ldots,k_m \in \IN$) such that
$\lambda_{k_1 \ldots k_m}$ is symmetric in $k_2,\ldots,k_m$ and such that

$$\Lambda J e_{k_1} = \underset{m}\to{\sum}\underset{k_2,\ldots,k_m}\to{\sum}
\lambda_{k_1 \ldots k_m} :x_{k_2}\cdots x_{k_m}: \quad , $$

\nind for arbitrary $k_1 \in \IN$. The $\lambda_{k_1 \ldots k_m}$ are
symmetric in the indices according to Lemma 3.1.1 and (3.1.6):
$$\align
\lambda_{k_1 k_2 k_3 \ldots k_m} &= \frac{2^m}{m!}\langle\Lambda Je_{k_1},
 :x_{k_2}x_{k_3}\cdots x_{k_m}: \rangle \\
&= \frac{2^m}{m!}\langle\Lambda Je_{k_2}, :x_{k_1}x_{k_3}\cdots x_{k_m}: 
\rangle \\
&= \lambda_{k_2 k_1 k_3 \ldots k_m} \quad .
\endalign $$
\nind The function defined by
$$G = \underset{m}\to{\sum}\underset{\{ 1,2\} \cap \{ k_1,\ldots,k_m\}\neq\emptyset}\to{\sum}
\frac{\lambda_{k_1 \ldots k_m}}{m} :x_{k_1}\cdots x_{k_m}: \quad , $$
\nind is an $L^2$-function, since by (3.1.1) one has
$$\align
\Vert G \Vert_2^2 &= 
\underset{m}\to{\sum}\underset{\{ 1,2\} \cap \{ k_1,\ldots,k_m\}\neq\emptyset}\to{\sum}
\frac{\lambda^2_{k_1 \ldots k_m}}{m^2}\frac{m!}{2^m} \\
&\leq \underset{m}\to{\sum}\underset{1 \in \{ k_1,\ldots,k_m\}}\to{\sum}
\frac{\lambda^2_{k_1 \ldots k_m}}{m^2}\frac{m!}{2^m} +
\underset{m}\to{\sum}\underset{2 \in \{ k_1,\ldots,k_m\}}\to{\sum}
\frac{\lambda^2_{k_1 \ldots k_m}}{m^2}\frac{m!}{2^m} \\
&\leq \underset{m}\to{\sum}\underset{l_2,\ldots,l_m}\to{\sum}m
\frac{\lambda^2_{1 l_2 \ldots l_m}}{m^2}\frac{m!}{2^m} +
 \underset{m}\to{\sum}\underset{l_2,\ldots,l_m}\to{\sum}m
\frac{\lambda^2_{2 l_2 \ldots l_m}}{m^2}\frac{m!}{2^m} \\
&= \underset{m}\to{\sum}\underset{l_2,\ldots,l_m}\to{\sum}
\lambda^2_{1 l_2 \ldots l_m}\frac{(m-1)!}{2^m} +
\underset{m}\to{\sum}\underset{l_2,\ldots,l_m}\to{\sum}
\lambda^2_{2 l_2 \ldots l_m}\frac{(m-1)!}{2^m} \\
&= \frac{1}{2}\Vert \underset{m}\to{\sum}\underset{l_2,\ldots,l_m}\to{\sum}
\lambda_{1 l_2 \ldots l_m} :x_{l_2} \cdots x_{l_m}: \Vert^2_2 \\
&\quad + \frac{1}{2}\Vert \underset{m}\to{\sum}\underset{l_2,\ldots,l_m}\to{\sum}
\lambda_{2 l_2 \ldots l_m} :x_{l_2} \cdots x_{l_m}: \Vert^2_2 \\
&= \frac{1}{2} \Vert\Lambda Je_1 \Vert_2^2 + \frac{1}{2}\Vert\Lambda Je_2 \Vert^2_2 < \infty \quad .
\endalign $$
\nind Then
$$\align
\frac{\partial}{\partial x_1}G &= \underset{m_0 \rightarrow\infty}\to{\lim}
\underset{m=1}\to{\overset{m_0}\to{\sum}}\,\underset{j=1}\to{\overset{m}\to{\sum}}\,\underset{k_i , i \neq j}\to{\sum}\frac{\lambda_{k_1 \ldots k_{j-1}1k_{j+1}\ldots k_m}}{m}:x_{k_1} \cdots x_{k_{j-1}}x_{k_{j+1}}\cdots x_{k_m}: \\
&= \underset{m}\to{\sum}\underset{l_2,\ldots,l_m}\to{\sum}
\lambda_{1 l_2 \ldots l_m} :x_{l_2} \cdots x_{l_m}: \\
&= \Lambda J e_1 \quad . 
\endalign $$
\nind But $\Lambda v_1$ is a constant function and 
$\langle \Lambda v_1,\Omega\rangle = 0$, so $\Lambda v_1 = 0$ and
$c_1 \frac{\partial}{\partial x_1}G = \Lambda f$. Proposition 3.1.3 then entails
$$\Phi_{\Lambda}(f) = e^{-iG}\Phi(f)e^{iG} \quad . $$
\nind Similarly, one has
$$\frac{\partial}{\partial x_2} G = \Lambda J e_2 \quad , $$
\nind and, since 
$$\align 
\frac{\partial}{\partial x(c_2 e_1 + c_3 e_2)}G &= 
c_2\frac{\partial}{\partial x_1}G + c_3\frac{\partial}{\partial x_2}G \\
&= c_2\Lambda Je_1 + c_3 \Lambda Je_2 \\
&= \Lambda g 
\endalign $$
\nind (recall that $\frac{\partial}{\partial x(c_2 e_1 + c_3 e_2)} =
\sqrt{2}(a(c_2 e_1 + c_3 e_2)) = \sqrt{2}(c_2 a(e_1) + c_3 a(e_2)) = 
c_2\frac{\partial}{\partial x_1} + c_3\frac{\partial}{\partial x_2}$ and
use $\Lambda v_2 = 0$, as well), Proposition 3.1.3 also implies
$$\Phi_{\Lambda}(g) = e^{-iG}\Phi(g)e^{iG} \quad . $$
\nind Hence, one has $\Phi_{\Lambda}(f+g) = e^{-iG}\Phi(f+g)e^{iG}$, which
implies
$$\align
e^{i\Phi_{\Lambda}(f)}e^{i\Phi_{\Lambda}(g)} &= 
e^{-iG}e^{i\Phi(f)}e^{i\Phi(g)}e^{iG} \\
&= e^{-iG}e^{-\frac{i}{2}\sigma(f,g)}e^{i\Phi(f+g)}e^{iG} \\
&= e^{-\frac{i}{2}\sigma(f,g)}e^{i\Phi_{\Lambda}(f+g)} \quad , 
\endalign $$
\nind and the proof is complete.
\hfill\qed\enddemo

     It follows from Theorem 3.1.5 that each $\Lambda \in \Ls_{CCR}$ is of the
form $\Lambda = \Lambda_l + \Lambda_q$, where $\Lambda_l \in \Ls_{CCR}$ is
linear, that is to say, the associated field operators 
$\Phi_{\Lambda_l}(f)$, $f \in D(\Lambda_l)$, are of the form 
$\Phi(g) + c\idty$, for suitable $g \in H$ and $c \in \RR$.  In other words,
the transformation $\Phi(f) \mapsto \Phi_{\Lambda_l}(f)$ is one of the 
inhomogeneous linear canonical transformations alluded to in Chapter II. It 
then follows that, with a suitable 
choice of linear $\Lambda_l$, the operator $\Lambda_q \in \Ls_{CCR}$ satisfies 
$P_0 \Lambda_q f = 0 = P_1 \Lambda_q f$,
for any $f \in D(\Lambda_q)$. The set of such operators $\Lambda_q$ will
be denoted by $\Ls^q_{CCR}$. The superscript $q$ is chosen because the degree
of such transformations is quadratic or higher. The structure of the linear 
elements of $\Ls_{CCR}$ will be discussed in Chapter IV. Here we shall 
consider the elements of $\Ls^q_{CCR}$. The following result generalizes 
Proposition 3.3.4 in \cite{22} from the quadratic case to this general setting. 

\proclaim{Proposition 3.1.7} Let $\Lambda \in \Ls^q_{CCR}$. Then there exists
a unique maximal extension $\Lambda_{max} \in \Ls^q_{CCR}$ of $\Lambda$.
\endproclaim

\demo{Proof} By Theorem 3.1.6, it may be assumed that $V \subset D(\Lambda)$,
and by suitably choosing the basis $\{ e_k \mid k \in \IN \}$ of $V$, it
may also be assumed that $\{ e_k, Je_k \mid k \in \IN \} \subset D(\Lambda)$.
If $f = f_1 + Jf_2 \in D(\Lambda)$ with $f_1,f_2 \in V$, then Lemma 3.1.1 and
Theorem 3.1.6 entail
$$\align
\Lambda f &= \underset{m,k_1,\ldots,k_m}\to{\sum}\frac{2^m}{m!}
\langle \Lambda Jf_2,:x_{k_1}x_{k_2}\cdots x_{k_m}:\rangle :x_{k_1}\cdots x_{k_m}: \\
&= \underset{m,k_1,\ldots,k_m}\to{\sum}\frac{2^m}{m!}
\langle \Lambda Je_{k_1},:x(f_2)x_{k_2}\cdots x_{k_m}:\rangle :x_{k_1}\cdots x_{k_m}: \quad .
\endalign $$
\nind If $\Lambda' \in \Ls_{CCR}^q$ is an extension of $\Lambda$ and
$f = f_1 + Jf_2 \in D(\Lambda')$, with $f_1,f_2 \in V$, then 
$$\align
\infty &> \Vert \Lambda' f \Vert^2 \\
&= \Vert \underset{m,k_1,\ldots,k_m}\to{\sum}\frac{2^m}{m!}
\langle \Lambda Je_{k_1},:x(f_2)x_{k_2}\cdots x_{k_m}:\rangle :x_{k_1}\cdots x_{k_m}: \Vert^2 \\
&\overset{(3.1.1)}\to{=} \underset{m,k_1,\ldots,k_m}\to{\sum}\frac{2^m}{m!}
\vert \langle \Lambda Je_{k_1},:x(f_2)x_{k_2}\cdots x_{k_m}:\rangle\vert^2 \quad .
\endalign $$
\nind Hence, define $\Lambda_{max}$ by
$$D(\Lambda_{max}) = 
\{ f + Jg \mid f,g \in V, \underset{m,k_1,\ldots,k_m}\to{\sum}\frac{2^m}{m!}
\vert \langle \Lambda Je_{k_1},:x(g)x_{k_2}\cdots x_{k_m}:\rangle\vert^2 < \infty\}$$
\nind and 
$$\Lambda_{max}f = 
\underset{m,k_1,\ldots,k_m}\to{\sum}\frac{2^m}{m!}
\langle \Lambda Je_{k_1},:x(f_2)x_{k_2}\cdots x_{k_m}:\rangle :x_{k_1}\cdots x_{k_m}: \quad , $$

\nind for $f = f_1 + Jf_2 \in D(\Lambda_{max})$, $f_1,f_2 \in V$. Since all
other assertions are now clear, it remains only to show that
$\Lambda_{max} \in \Ls_{CCR}$. But for
$f_1,\ldots,f_m \in JD(\Lambda_{max}) \cap V$ (with $f_l = \sum_k c_{lk}e_k$),
one sees
$$\align
\quad &\langle\Lambda_{max}Jf_1,:x(f_2)\cdots x(f_m):\rangle \\
&= \underset{k_2,\ldots,k_m}\to{\sum} c_{2k_2}\cdots c_{mk_m}\langle
\Lambda_{max}Jf_1,:x(e_{k_2})\cdots x(e_{k_m}):\rangle \\
&= \underset{k_2,\ldots,k_m}\to{\sum} c_{2k_2}\cdots c_{mk_m}\langle
\Lambda Je_{k_m},:x(f_1)x(e_{k_2})\cdots x(e_{k_{m-1}}):\rangle \\
&= \underset{k_m}\to{\sum}c_{mk_m}
\langle\Lambda Je_{k_m},:x(f_1)x(f_2)\cdots x(f_{m-1}):\rangle \\
&= \underset{k_m}\to{\sum}c_{mk_m}
\langle\Lambda Je_{k_m},:x(f_2)x(f_1)\cdots x(f_{m-1}):\rangle \\
&= \langle\Lambda_{max}Jf_2,:x(f_1)x(f_3)\cdots x(f_m):\rangle \quad .
\endalign $$
\nind Theorem 3.1.6 then implies $\Lambda_{max} \in \Ls_{CCR}$.
\hfill\qed\enddemo

     We then can use this result to show, as in Section 3.3 of \cite{22} for 
the quadratic case, that any $\Lambda \in \Ls^q_{CCR}$ has a particular
form, which leads to a convenient ``standard'' form for the corresponding
field operators $\Phi_{\Lambda}(f)$. 

\proclaim{Proposition 3.1.8} For each $\Lambda \in \Ls^q_{CCR}$ there exist
an orthonormal basis $\{ e_k \mid k \in \IN\}$ of $V$ and real numbers
$\lambda_{k k_1 \ldots k_m}$ totally symmetric in the indices such that
all of the following conditions are satisfied:

$$\underset{m,k_1,\ldots,k_m}\to{\sum}\lambda^2_{k k_1 \ldots k_m}
\frac{m!}{2^m} < \infty \quad , $$

\nind for any $k \in \IN$; moreover, if $\Lambda' \in \Ls^q_{CCR}$ is defined
on the linear span of $\{ e_k , Je_k \mid k \in \IN\}$ by $\Lambda' e_k = 0$
and 
$\Lambda' Je_k = \sum_{m,k_1,\ldots,k_m}\lambda_{k k_1 \ldots k_m}:x_{k_1}\cdots
x_{k_m}:$, then $\Lambda \subset \Lambda'_{max}$.
\endproclaim

\demo{Proof} Once again, it may be assumed that $V \subset D(\Lambda)$ and
$\{ Je_k \mid k \in \IN\} \subset D(\Lambda)$. Set

$$\lambda_{k k_1 \ldots k_m} = \frac{2^m}{m!}\langle\Lambda Je_k,:x_{k_1}\cdots
x_{k_m}:\rangle \quad . $$

\nind Then, by Lemma 3.1.1, one finds

$$\align
\infty &> \Vert\Lambda Je_k \Vert^2 \\
&= \Vert \underset{m,k_1,\ldots,k_m}\to{\sum}\lambda_{k k_1 \ldots k_m}
:x_{k_1}\cdots x_{k_m}: \Vert^2 \\
&= \underset{m,k_1,\ldots,k_m}\to{\sum}\lambda^2_{k k_1 \ldots k_m}\frac{m!}{2^m}
\quad .
\endalign $$

\nind Proposition 3.1.7 then completes the proof.
\hfill\qed\enddemo

     Since $\Lambda_{max}$ uniquely exists, one may consider 
$\Lambda \in \Ls_{CCR}^q$ as being defined on a symplectic orthonormal basis
$\{ e_k,Je_k \}_{k \in \IN}$ such that $\Lambda e_k = 0$ and
$\Lambda Je_k = \sum_{m,k_1,\ldots,k_m}\lambda_{k k_1 \ldots k_m}:x_{k_1}\cdots
x_{k_m}:$. Then one has

$$q_k \equiv \Phi(e_k) \mapsto \Phi_{\Lambda}(e_k) = \Phi(e_k) = q_k \quad , $$

\nind and

$$\align
p_k \equiv \Phi(Je_k) \mapsto \Phi_{\Lambda}(Je_k) &=
\overline{p_k + \sum_{m,k_1,\ldots,k_m}\lambda_{k k_1 \ldots k_m}:x_{k_1}\cdots
x_{k_m}:\idty} \\
&= \overline{p_k + \sum_{m,k_1,\ldots,k_m}\lambda_{k k_1 \ldots k_m}:q_{k_1}\cdots q_{k_m}:} \quad .
\endalign $$

     In short, the standard form of a canonical transformation of arbitrary 
degree is that given in equation (1.1). This standard form, along with being 
physically more transparent, was useful in the special case of quadratic
transformations in \cite{22} to establish results concerning the unitary 
equivalence of such representations with the Fock representation. Though we do 
not prove such results here for transformations of arbitrary degree, we 
{\it shall} give necessary and sufficient conditions for unitary, resp.
quasi-, equivalence between Fock, coherent, and quasifree representations and 
representations of finite degree in Chapters IV and V. \par

\bigpagebreak

\proclaim{3.2 Irreducibility of the Representation}
\endproclaim

     In this section we shall show that for bounded $\Lambda \in \Ls_{CCR}$, 
the corresponding representation $\pi_{\Lambda}$ of the CCR is irreducible.
If $\Lambda$ is unbounded, then it can occur that $\pi_{\Lambda}$ is reducible,
as we shall explain. We begin with a technical lemma concerning the 
closability and continuity properties of $\Lambda \in \Ls_{CCR}$ such that 
$P_0 \Lambda$ is the zero operator, {\it i.e.} such that the range of 
$\Lambda$ is orthogonal to the vacuum vector $\Omega$. Note that this is true 
of each $\Lambda \in \Ls_{CCR}$, up to a coherent transformation, {\it i.e.}
a transformation of degree zero.

\proclaim{Lemma 3.2.1} Let $\Lambda \in \Ls_{CCR}$ with 
$P_0 \Lambda \subset 0$.
Then $\Lambda$ is closable and $\overline{\Lambda} \in \Ls_{CCR}$.
Furthermore, for any sequence $\{f_n \}_{n \in \IN} \subset D(\Lambda)$
such that $f_n \rightarrow f \in D(\overline{\Lambda})$ and 
$\Lambda f_n \rightarrow \overline{\Lambda}f$ as $n \rightarrow \infty$,
then the operators $\{e^{i\Phi_{\Lambda}(f_n)}\}_{n \in \IN}$ converge strongly
to $e^{i\Phi_{\overline{\Lambda}}(f)}$ as $n \rightarrow \infty$.
\endproclaim

\demo{Proof} It will first be shown that such $\Lambda$ are closable.
Once again, it may be assumed that $V \subset D(\Lambda)$ and
$\{ Je_k \mid k \in \IN\} \subset D(\Lambda)$. Let $\{ g_n\}$ in $D(\Lambda)$ 
be a sequence such that $g_n = h_n + Jh'_n$, with
$h_n,h'_n \in V$, $g_n \rightarrow 0$ and $\Lambda g_n \rightarrow F$,
for some $F \in \L2Q$. Then for arbitrary $m,k_1,\ldots,k_m \in \IN$, Theorem
3.1.6 implies

$$\align
\langle\Lambda g_n,:x_{k_1}\cdots x_{k_m}:\rangle &= 
\langle\Lambda Jh'_n,:x_{k_1}x_{k_2}\cdots x_{k_m}:\rangle  \\
&= \langle\Lambda Je_{k_1},:x(h'_n)x_{k_2}\cdots x_{k_m}:\rangle  \\
&\rightarrow 0 \quad , 
\endalign $$

\nind as $n \rightarrow \infty$. Hence, $F = 0$ and $\Lambda$ is closable.
Since $\overline{\Lambda}$ fulfills (3.1.4), Theorem 3.1.5 entails 
$\overline{\Lambda} \in \Ls_{CCR}$. \par
     In addressing the final assertion in the lemma, one may assume that
$V \subset D(\Lambda)$, $\Lambda = \overline{\Lambda}$, $f = Je_1 + v$,
with $v \in V$, and

$$\Lambda f = \underset{m}\to{\sum} \underset{l=0}\to{\overset{\infty}\to{\sum}}
\quad \underset{k_1,\ldots,k_m \geq 2}\to{\sum} c_{lk_1 \ldots k_m}
:x^l_1x_{k_1}\cdots x_{k_m}: \quad ,$$

\nind for suitable $c_{lk_1 \ldots k_m}$, symmetric in the indices
$k_1 \ldots k_m$. Set

$$G_n = \underset{m+l \leq n}\to{\underset{2\leq k_1,\ldots,k_m \leq n}\to{\sum}}
c_{lk_1 \ldots k_m}\frac{1}{l+1} :x^{l+1}_1x_{k_1}\cdots x_{k_m}:
$$

\nind and

$$G = \underset{m,l, k_1,\ldots,k_m}\to{\sum}
c_{lk_1 \ldots k_m}\frac{1}{l+1} :x^{l+1}_1x_{k_1}\cdots x_{k_m}: \quad .
$$

\nind Note that the proof of Proposition 3.1.3 implies that $G \in \L2Q$. Then one
has

$$\align
\Phi_{\Lambda}(f_n)e^{-iG_m}\varphi &= \Phi(f_n)e^{-iG_m}\varphi +
\Lambda f_n e^{-iG_m}\varphi \\
&\rightarrow \Phi(f)e^{-iG_m}\varphi + \Lambda f e^{-iG_m}\varphi \\
&= \Phi_{\Lambda}(f)e^{-iG_m}\varphi \quad ,
\endalign $$

\nind for $\varphi \in \Gs$. But the set
$\{ e^{-iG_m}\varphi \mid m \in \IN, \varphi \in \Gs \}$ is contained in the
domain of the strong graph limit of the sequence $\{ \Phi_{\Lambda}(f_n)\}$,
and this strong graph limit is a symmetric and closed operator (see, 
{\it e.g.}, Theorem VIII.27 in \cite{24}). Furthermore, this strong
graph limit acts upon the elements of this set as $\Phi_{\Lambda}(f)$. As
in the proof of Proposition 3.1.3, one may conclude that $e^{-iG_m}\varphi$
converges to $e^{-iG}\varphi$ as $m \rightarrow \infty$ and, since
$\frac{\partial}{\partial x_1}G_m$ converges to $\Lambda f$ in the 
$L^2$-norm, 

$$\Phi_{\Lambda}(f)e^{-iG_m}\varphi = e^{-iG_m}(\Phi(f) + \Lambda f \idty - 
\frac{\partial}{\partial x_1}G_m \idty)\varphi \rightarrow
e^{-iG}\Phi(f)\varphi \quad . $$

\nind In other words, the set $e^{-iG}\Gs$ is contained in the domain of
the strong graph limit of the sequence $\{ \Phi_{\Lambda}(f_n)\}$, and it
acts upon this set as $\Phi_{\Lambda}(f) = e^{-iG}\Phi(f)e^{iG}$. But
$\Gs$ is a core for $\Phi(f)$, so it follows that $\Phi_{\Lambda}(f)$ is, in
fact, the strong graph limit of $\{ \Phi_{\Lambda}(f_n)\}$. The final assertion
of the lemma then follows from Theorems VIII.21 and VIII.26 in 
\cite{24}.
\hfill\qed\enddemo

     We now provide a sufficient condition on $\Lambda$ which entails
that $\pi_{\Lambda}$ is irreducible.

\proclaim{Theorem 3.2.2} Let $\Lambda \in \Ls_{CCR}$. If a dense subset of
$V$ is contained in $D(\overline{\Lambda - P_0 \Lambda})$, then
$\pi_{\Lambda}$ is irreducible. In particular, $\pi_{\Lambda}$ is irreducible
for bounded $\Lambda - P_0 \Lambda$, resp. for bounded $\Lambda$.
\endproclaim

\demo{Proof} Let $T : \L2Q \rightarrow \L2Q$ be bounded and assume

$$[T,e^{i\Phi_{\Lambda}(g)}] = 0 \quad , $$

\nind for any $g \in D(\Lambda)$. Then 

$$[T,e^{i\Phi(g)}] = 0 \quad , \tag{3.2.1}$$

\nind for $g$ in a dense subset of $V$. But Lemma 3.2.1 implies that equation
(3.2.1) holds for all $g \in V$. Since $\{ e^{ix(f)} \mid f \in V \}$ generates
the maximally abelian von Neumann algebra $L^{\infty}(Q,d\mu)$, $T$ may be
identified with an element of $L^{\infty}(Q,d\mu)$ (or, more accurately,
with the corresponding multiplication operator). \par
     Now, Proposition 3.1.3 implies that for any $g \in D(\Lambda) \setminus V$
there exists a $G \in \L2Q$ such that 

$$e^{i\Phi_{\Lambda}(g)} = e^{-iG}e^{i\Phi(g)}e^{iG} \quad . $$

\nind But $[T,e^{iG}] = 0$ entails equation (3.2.1) for 
$g \in D(\Lambda) \setminus V$ and, therefore, by Lemma 3.2.1, also for any
$g \in H$. Since any Fock representation is irreducible, it follows that
$T$ is a multiple of the identity. 
\hfill\qed\enddemo

     We wish to show that there do exist $\Lambda \in \Ls_{CCR}$ such that
$\pi_{\Lambda}$ is reducible. To set this up properly, we first prove the
following lemma.

\proclaim{Lemma 3.2.3} If $\Ks$ is a Hilbert space and 
$T : \Ks \supset D(T) \rightarrow \Ks$ is an unbounded densely defined operator,
then there exists a densely defined $S \subset T$ such that 
$\overline{R(S)}\neq\Ks$.
\endproclaim

\demo{Proof} The equality $D(T^*) = \Ks$ would imply the boundedness of
$T^*$ and therefore of $T$. Thus, there exists an $f \in \Ks \setminus D(T^*)$,
with which one may define the operator $S$ as the restriction of $T$ to

$$ D(S) = \{ g \in D(T) \mid \langle f,Tg \rangle = 0 \} \quad . $$

\nind $f \not\in D(T^*)$ entails the existence of a unit vector $f_n \in D(T)$
such that

$$\underset{n\rightarrow\infty}\to{\lim} \vert\langle f,Tf_n \rangle\vert \rightarrow \infty \quad . $$

\nind It may be assumed that $f_n \not\in D(S)$. Let $g \in D(T)$ be
arbitrary. Then

$$g - \frac{\langle f,Tg \rangle}{\langle f,Tf_n \rangle}f_n \in D(S) \quad
\text{and} \quad g - \frac{\langle f,Tg \rangle}{\langle f,Tf_n \rangle}f_n 
\rightarrow g \quad , $$

\nind as $n \rightarrow \infty$. Hence, $g \in \overline{D(S)}$. Since $D(T)$
is dense in $\Ks$, this establishes that $\overline{D(S)} = \Ks$.
\hfill\qed\enddemo

     We can now show that there exists a linear $\Lambda \in \Ls_{CCR}$ such 
that $\pi_{\Lambda}$ is reducible. In fact, to each linear element $\Lambda$ of
$\Ls_{CCR}$ there corresponds a symplectic transformation, which is 
unbounded if $\Lambda$ is unbounded (see \cite{26}). According
to Lemma 3.2.3, we may restrict the domain of any unbounded symplectic
transformation to a set which is still dense in $H$ in such a manner that
the range of the restriction is not dense in $H$. However, the representation
induced by a symplectic transformation is irreducible if and only if the
range of the symplectic operator is dense in $H$, as we shall see in the
next chapter (Lemma 4.1).\par
     We also wish to point out that Corollary 4.1.2 in \cite{22} is false
as stated; in particular, the claim of irreducibility does not follow. As
discussed in \cite{11}, the argument sketch given in \cite{22} tacitly 
assumed that $\overline{\Lambda}$ is defined on a proper standard basis,
which certainly follows if $\overline{\Lambda} = \Lambda_{max}$, and hence
also if $\Lambda$ is bounded, but which is not true in general. Florig also 
provides an example of a {\it reducible} quadratic representation (see 
Section 2.3 in \cite{11}).

\bigpagebreak

\heading IV. Quasifree States and Linear Canonical Transformations \endheading

     In this chapter we shall restrict our attention to quasifree states on
$\As(H)$ and the associated representations. The notion of quasifree state
was introduced by D.W. Robinson \cite{27} in his study of the ground state
of the Bose gas. It was shown in \cite{18} that such (pure) states can be
obtained by Bogoliubov transformations of a Fock state, hence making it clear
that the class of representations (commonly called symplectic representations)
studied by Segal \cite{32} and Shale \cite{33}, among others, 
essentially coincided with the quasifree representations.
We wish to show that the GNS representation of any pure quasifree state is
unitarily equivalent to one of the representations $\pi_{\Lambda}$ constructed
in the previous chapter, for a suitable choice of $V \subset H$ and a linear
$\Lambda \in \Ls_{CCR}$. The polynomial representations constructed globally 
under certain boundedness restrictions in \cite{30} and the quadratic 
representations of \cite{22} are clearly included among the representations of 
finite degree (special cases of the class constructed in Chapter III) 
discussed in more detail in Chapter V. Since the coherent representations are 
special cases of pure quasifree representations (see Proposition 4.4) and are
therefore also subsumed in the class of representations presented in Chapter 
III, we see that our methods serve to unify the approaches to these various 
classes of representations, as well as to extend them to arbitrary degree. \par
     We recall that if $\omega_J$ is a Fock state on $\As(H)$ with associated 
representation $(\Ks, \pi_J )$, then a coherent state $\omega_l$ is given by

$$\omega_l(W(f)) \equiv \omega_J(W(f))e^{i\, l(f)}\quad , \quad f \in H \quad .$$

\nind The GNS representation of $\As(H)$ corresponding to $\omega_l$ is given 
on $\Ks$ by 
$$\pi_l(W(f)) \equiv e^{i\, l(f)}\pi_J(W(f)) \quad , \quad f \in H \quad . $$ 
From Theorem 3.1 of \cite{26} (but see also \cite{34}\cite{4}), it follows 
that the representations $\pi_l$ and $\pi_J$ are unitarily equivalent if and 
only if the map $l : H \rightarrow \RR$ is bounded. And if a quasifree 
representation is obtained from a given Fock representation of $\As(H)$ by
$$\Phi_T(f) \equiv \Phi(Tf) = \Phi(f) + \Phi((T-\idty)f) \quad , $$ 
using a symplectic operator $T$, it is known \cite{33} that the representations
$\pi_J$ and $\pi_T$ are unitarily equivalent if and only if the operator
$\idty - \vert T \vert $ is Hilbert-Schmidt. (See Theorem 3.2 in \cite{26} for 
a basis-dependent formulation of this result.) \par
     Given a dense subspace $H_0 \subset H$, we shall show that all the
quasifree states on $\As(H_0)$ can be obtained from the Fock state $\omega_J$ 
by symplectic transformations $\Lambda \in \Ls_{CCR}$. To begin, we consider
pure quasifree states. It is known that, in our terminology, a pure
quasifree state is, up to a coherent transformation, a Fock state
\cite{18}. In order to be more precise, we need to introduce some
notation. \par
     Let $\omega'_F$ be a Fock state on $\As(H_0)$, so there exists an
associated scalar product $s'$ on $H_0$ \cite{17}, with respect to which 
the completion of $H_0$ will be denoted by $H'$ (and the induced scalar product
on $H'$ will again be called $s'$). The scalar product $s'$ is such that
the symplectic form $\sigma$ determining $\As(H)$ is continuous with respect
to $s'$, when restricted to $H_0$ \cite{17}. Hence, the restriction of 
$\sigma$ to $H_0$ extends uniquely to a nondegenerate symplectic bilinear form 
$\sigma'$ on $H'$. Moreover, there exists an operator $J' : H' \rightarrow H'$
which induces a complex structure on $H'$ \cite{14}, so that, in particular,

$$s'(f,g) = -\sigma'(J'f,g) \quad , $$

\nind for all $f,g \in H'$. Since $H_0$ is dense in $H$, resp. $H'$,
with respect to $s$, resp. $s'$, and $\sigma$ and $\sigma'$ are
nondegenerate, we may assume $f = f' \in H \cap H'$, whenever

$$\sigma(f,h) = \sigma'(f',h) \quad f \in H \, , \, f' \in H' \, , \, \forall
h \in H_0 \quad . \tag{4.1}$$

\nind There is no loss of generality, since we do not assume that
$\sigma(f_1,f_2) = \sigma'(f_1,f_2)$ for $f_1,f_2 \in H \cap H'$. \par
     The existence of a symplectic $T : H \rightarrow H$ with
$s'(f,f) = s(Tf,Tf)$, for all $f \in H$, was already proven for
$H = H_0 = H'$ in \cite{18}. We shall need to generalize this result.
First, we characterize symplectic maps and irreducible symplectic
representations. 

\proclaim{Lemma 4.1} An operator $T : H \supset D(T) \rightarrow H$ is
symplectic (with respect to $\sigma$), {\it i.e.} 
$\sigma(Tf,Tg) = \sigma(f,g)$ for all $f,g \in D(T)$, if and only if

$$-JT^{-1}J \subset T^*$$

\nind (it is not assumed here that $T^{-1}$, resp. $T^*$, is necessarily
densely defined). A self-adjoint operator $T$ is symplectic if and only if
$-JT^{-1}J = T$. For symplectic $T$, the representation $\pi_T$ defined by

$$\pi_T(W(f)) = e^{i\Phi(Tf)} \quad , \quad f \in D(T) \quad , $$

\nind is irreducible if and only if $\overline{R(T)} = H$.
\endproclaim

\demo{Proof} Let $T$ be symplectic and $g \in D(T)$. Then 
one has for $f \in D(JT^{-1}J)$

$$\align
s(-JT^{-1}Jf,g) &= \sigma(-T^{-1}Jf,g) = -\sigma(Jf,Tg) \\
&= s(f,Tg) = s(T^* f,g) \quad . 
\endalign $$

\nind Hence, $-JT^{-1}J \subset T^*$. The converse follows from the equalities
$$\align
\sigma(f,g) &= s(JT^{-1}Tf,g) = s(T^* JTf,g) \\
&= s(JTf,Tg) = \sigma(Tf,Tg) \quad , 
\endalign $$

\nind for all $f,g \in D(T)$. If $T$ is symplectic and self-adjoint, then 
$T^{-1}$ is densely defined
($T$'s null space is trivial), so, by the above result, $T$ is a self-adjoint 
extension of $-JT^{-1}J = -J(T^*)^{-1}J = (-JT^{-1}J)^*$, and thus
$T = -JT^{-1}J$. \par
     Turning to the characterization of irreducible symplectic representations,
if $g \in R(T)^{\bot}$, then 
$e^{i\Phi(Jg)} \in \pi_T(\As(R(T)))'$, so that $\pi_T$ is reducible if
$\overline{R(T)} \neq H$. Assume now that $\overline{R(T)} = H$. If
the sequence $\{ f_n \}_{n \in \IN} \subset H$ converges to $f \in H$, then
$\{e^{i\Phi(f_n)}\}_{n\in\IN}$ converges to $e^{i\Phi(f)}$ strongly (use Lemma 
3.1.1 with $\Lambda = 0$). Thus, an element of the commutant of 
$\pi_T(\As(R(T)))$ must commute with the elements of $\pi_J(\As(H))$, which is 
itself a Fock representation and, hence, irreducible.
\hfill\qed\enddemo 

     With this in hand, we can now generalize the mentioned result of
Manuceau and Verbeure.

\proclaim{Proposition 4.2} Given the above-established notation, there exists 
a subspace $H_1 \supset H_0$ of
$H \cap H'$ such that the following conditions are fulfilled. If one defines
an operator $K$ by $K \subset J'$ and 
$D(K) = \{ f \in H_1 \mid J'f \in H_1 \}$, then 
$-JK : H \supset D(-JK) \rightarrow H$ is a symplectic (with respect to $\sigma$)
positive self-adjoint (with respect to $s$) operator, and $T = (-JK)^{1/2}$
is a symplectic transformation with $D(T) = H_1$ and

$$s'(f,g) = s(Tf,Tg) \quad , \quad \forall f,g \in H_1 \quad .$$

\endproclaim

\demo{Proof} By using Zorn's Lemma, it is easy to show that there exists a 
subspace $H_1 \supset H_0$ of $H \cap H'$ such that $H_1$ is maximal with
the property

$$\sigma(f,g) = \sigma'(f,g) \quad , \quad \forall f,g \in H_1 \quad . 
\tag{4.2} $$

\nind Consider the restriction $s'_{H_1}$ of the positive quadratic form
$s'$ determined by the form core $Q(s'_{H_1}) = H_1$. Because of the
aforesaid maximality, $H_1$ is closed with respect to the norm

$$\Vert f \Vert = \sqrt{s'(f,f) + s(f,f)} \quad , \quad f \in H \cap H' \quad .
$$

\nind The quadratic form $s'_{H_1}$ determines a self-adjoint operator
$A : H \supset D(A) \rightarrow H$ (use, {\it e.g.} Theorem VIII.15 in 
\cite{24}),
and the closure of $D(A)$ with respect to the above norm $\Vert\cdot\Vert$ is 
$H_1$. Hence, 

$$\sigma(f,JAg) = s(f,Ag) = s'_{H_1}(f,g) = \sigma'(f,J'g) \quad , \quad
f,g \in D(A) \subset H_1 $$
 
\nind is also true for $f \in H_1$. But the equality
$\sigma(f,JAg) = \sigma'(f,J'g)$, for any $f \in H_1 \supset H_0$ entails
the equality $JAg = J'g \in H \cap H'$ (see (4.1)) and, thus, by the maximality
of $H_1$, the equality $JAg = J'g \in H_1$, for $g \in D(A)$. According to the
definition of $K$, one has $Kg = J'g$, for $g \in D(K)$, hence
$Ag = -JJ'g = -JKg$, for $g \in D(A)$. From (4.2) one sees

$$\align
s(f,-JKg) &= \sigma(f,Kg) = \sigma'(f,Kg) \\
&= -\sigma'(Kf,g) = -\sigma(Kf,g) = s(-JKf,g) \quad , 
\endalign $$

\nind for $f,g \in D(-JK) \subset H_1$ (so $Kf,Kg \in H_1$), thus the operator 
$-JK \supset A$ is symmetric. Hence, $-JK = A$ is positive and self-adjoint,
and one can define the positive self-adjoint operator $T = (-JK)^{1/2}$.
The equality $s'(f,g) = s(Tf,Tg)$, which holds for all $f,g \in D(A)$, is
therefore still true for $f,g \in D(T) = H_1$, which is the closure of
$D(A)$ with respect to the norm $\Vert\cdot\Vert$. \par
     The operator $-JK$ is symplectic with respect to $\sigma$, since

$$\sigma(-JKf,-JKg) = \sigma(Kf,Kg) = \sigma'(Kf,Kg) = \sigma'(f,g) = 
\sigma(f,g) \quad , $$

\nind for $f,g \in D(-JK) \subset H_1$ (so $Kf,Kg \in H_1$). \par
     It remains to prove that $T$ is symplectic. From \cite{33} and
Lemma 4.1, one can decompose $H$ and $-JK$ as follows:

$$H = U \oplus JU \quad , \quad -JK = L \oplus -JL^{-1}J \quad ,$$

\nind with $L : U \rightarrow U$ self-adjoint and $0 \leq L \leq \idty$.
Therefore, $T = L^{1/2} \oplus -JL^{-1/2}J = -JT^{-1}J$ is symplectic,
by Lemma 4.1.
\hfill\qed\enddemo

     This permits us to characterize pure quasifree states. (If the proof
is not yet clear, then read the first few lines of the proof of Proposition
4.4.)

\proclaim{Corollary 4.3} Let $H_0$ be a dense subspace of $H$. Each pure 
quasifree state $\omega$ on $\As(H_0)$ has a characteristic function of the
form

$$\omega(W(f)) = e^{il(f)-\frac{s(Tf,Tf)}{4}} \quad , \quad f \in H_0 \quad , 
$$

\nind for some linear form $l : H_0 \rightarrow \RR$ and a symplectic positive
self-adjoint operator $T : H_0 \subset D(T) \rightarrow H$.
\endproclaim

     With this result, we can characterize general quasifree states.

\proclaim{Proposition 4.4} Let $H_0$ be a dense subspace of $H$. Each 
quasifree state $\omega$ on $\As(H_0)$ has a characteristic function of the
form

$$\omega(W(f)) = e^{il(f)-\frac{s(Tf,Tf)}{4}} \quad , \quad f \in H_0 \quad , 
$$

\nind for some linear form $l : H_0 \rightarrow \RR$ and a symplectic operator 
$T : H_0 \subset D(T) \rightarrow H$.
\endproclaim

\demo{Proof} By \cite{18}, $\omega$ has a characteristic function of the
form

$$\omega(W(f)) = e^{il(f)-\frac{s'(f,f)}{4}} \quad , \quad f \in H_0 \quad , 
$$

\nind for some scalar product $s'$ on a Hilbert space $M \supset H_0$. The
symplectic form $\sigma$ can be continuously (with respect to $s'$) extended
to a bilinear form $\sigma'$ on $M$ (see inequality (2) in \cite{18}). Let
$P : M \rightarrow M$ be the orthogonal projection onto the closure of
$\{ f \in M \mid \sigma(f,g) = 0 \, , \, \forall g \in M \}$ and set 
$Q = \idty - P$. The inequality (see, once again, (2) in \cite{18})

$$\vert\sigma'(f,g)\vert^2 = \vert\sigma'(Qf,Qg)\vert^2 \leq
s'(Qf,Qf)s'(Qg,Qg) \quad , $$

\nind for all $f,g \in M$, implies that one can define a state $\omega'$ on
$\As(H_0)$ by

$$\omega'(W(f)) = e^{-s'(Qf,Qf)/4} \quad , $$

\nind for all $f \in H_0$, by Proposition 10 in \cite{18}. According to Proposition 11
in the same paper, the state $\omega'$ is also primary, since the restriction 
of $\sigma'$ to $QM \times QM$ is nondegenerate. The discussion in Section IV 
of \cite{18} also implies the existence of a scalar product $s_0$ on $H_0$ 
associated with a pure state on $\As(H_0)$ such that

$$s_0(f,f) \leq s'(Qf,Qf) \quad , $$

\nind for all $f \in H_0$. Thus, there exists a symplectic operator
$S : H_0 \rightarrow H$ such that

$$s(Sf,Sf) = s_0(f,f) \leq s'(Qf,Qf) \leq s'(f,f) \quad , $$

\nind for all $f \in H_0$, using Corollary 4.3. Furthermore, from Theorem
VIII.15 in \cite{24} one has

$$s'(f,g) = s(A^{\frac{1}{2}}f,A^{\frac{1}{2}}g) \quad , \forall f,g \in
D(A^{\frac{1}{2}}) \supset H_0 \quad , $$

\nind for a suitable self-adjoint $A$. But then
$s(A^{\frac{1}{2}}\cdot,A^{\frac{1}{2}}\cdot) - s(S\cdot,S\cdot)$
is a positive quadratic form on $H_0$, so, by appealing once again to
Theorem VIII.15 in \cite{24}, there exists a positive self-adjoint 
operator $B : H \supset D(B) \rightarrow H$ such that

$$ s(Bf,Bf) = s(A^{\frac{1}{2}}f,A^{\frac{1}{2}}f) - s(Sf,Sf) \quad , 
\tag{4.3}$$

\nind for all $f \in H_0$.  \par
     Now define isometries (with respect to $s$) $U,V : H \rightarrow H$ by

$$Ue_k = e_{3k} \, , \, UJe_k = Je_{3k} \quad \text{and} \quad 
Ve_k = e_{3k+1} \, , \, VJe_k = Je_{3k+2} \quad , k \in \IN \quad , $$

\nind and set $T = US + VB$. Then the equalities (by definition, $U$ commutes
with $J$)

$$\align
\sigma(Tf,Tg) &= \sigma((US + VB)f,(US + VB)g) = \sigma(USf,USg) \\
  &= \sigma(Sf,Sg) = \sigma(f,g) \quad ,
\endalign $$

\nind for all $f,g \in D(T)$, entail that $T$ is symplectic. The claim
then follows after noting that (4.3) implies

$$s(Tf,Tf) = s(Sf,Sf) + s(Bf,Bf) = s(A^{\frac{1}{2}}f,A^{\frac{1}{2}}f) 
= s'(f,f) \quad , $$

\nind for all $f \in H_0$.
\hfill\qed\enddemo

     It is now also clear that coherent representations are special cases of
pure quasifree representations. This permits us to prove the result announced 
at the beginning of the chapter. 

\proclaim{Theorem 4.5} Let $\pi$ be the GNS-representation associated to 
a pure quasifree state on $\As(H_0)$, where $H_0$ is a dense subspace of $H$.
For a suitable choice of $V \subset H$, there exists a linear 
$\Lambda \in \Ls_{CCR}$ such that $\pi$ is unitarily equivalent to
$\pi_{\Lambda}$. 
\endproclaim

\demo{Proof} From the above discussion, it may be assumed that there exist
a positive self-adjoint operator $T : H \supset D(T) \rightarrow H$ with
$H_0 \subset D(T)$ and a linear form $l : H_0 \rightarrow \RR$ such that

$$\pi(W(f)) = e^{i\Phi(Tf)+il(f)} \quad , $$

\nind for all $f \in H_0$. From the proof of Proposition 4.2 it is clear that
$H$ and $T$ decompose as

$$H = U \oplus JU \quad , \quad T = A \oplus -JA^{-1}J \quad , $$

\nind with $A : U \rightarrow U$ self-adjoint and satisfying $0 \leq A \leq \idty$.
Set

$$U_1 = \{(\idty + JA)\varphi \mid \varphi \in U \} \subset H \quad .$$

\nind It is easy to see that $JU_1 \subset U_1^{\bot}$. With $\varphi \in U$,
the inclusions

$$\varphi = (\idty + JA)\frac{1}{\idty + A^2}\varphi - 
J(\idty + JA)\frac{A}{\idty + A^2}\varphi \, \in \, U_1 \oplus JU_1 $$

\nind and 

$$J\varphi \in J(U_1 \oplus JU_1) = U_1 \oplus JU_1 $$

\nind imply $H = U_1 \oplus JU_1$. The restriction of $T$ to $U_1$ is then
an isometry, since 
$$T(\idty + JA)\varphi = (A + (-JA^{-1}J)JA)\varphi = J(-JA + \idty)\varphi \quad , $$

\nind for all $\varphi \in U$, and 
$$\align
\Vert T(\idty + JA)\varphi \Vert^2 &= \Vert J(-JA + \idty)\varphi\Vert^2 = 
\Vert JA\varphi\Vert^2 + \Vert \varphi \Vert^2 \\
&= \Vert (\idty + JA)\varphi \Vert^2 \quad . 
\endalign $$

    One can similarly prove that 
$H = \overline{TU_1} \oplus J\overline{TU_1}$, so there exist a unitary
(considering $\Hs$ instead of $H$) $W : H \rightarrow H$ such that $WT$ is the
identity on $U_1$. The equalities

$$0 = \sigma(WTf,WTg) - \sigma(f,g) = \sigma(f,(WT - \idty)g) \quad , $$

\nind for $f \in U_1$, $g \in D(T)$, entail $R(WT - \idty) \subset U_1$. Choose
now $V = U_1$ and define $\Lambda : H_0 \rightarrow \L2Q$ by
$\Lambda f = l(f) + x((WT-\idty)f)$ for any $f \in H_0$. Then
$\Lambda \in \Ls_{CCR}$ and $\Phi_{\Lambda}(f) = \Phi(WTf) + l(f)\idty$.
According to \cite{33}, since $\vert W \vert - \idty = 0$ is a 
Hilbert-Schmidt operator, there exists a unitary $L$ such that 
$L\Phi(f)L^* = \Phi(Wf)$, for all $f \in H$. In particular, one has
$L\Phi(Tf)L^* = \Phi(WTf)$, for all $f \in D(T)$. It is therefore clear that 
the representation $\pi_{\Lambda}$ corresponding to $\Phi_{\Lambda}(f)$ 
is unitarily equivalent to the representation $\pi$ given by

$$\pi(W(f)) = e^{i\Phi(Tf)+il(f)} \quad , \quad f \in H_0 \quad . $$

\hfill\qed\enddemo
 
     To close this chapter, we give a characterization of our linear canonical 
transformations. Recall that $P : H \rightarrow H$ is the orthogonal projection 
onto the subspace $JV$.

\proclaim{Proposition 4.6} To every linear $\Lambda \in \Ls_{CCR}$ there 
corresponds a linear form $l : D(\Lambda) \rightarrow \RR$ and a symmetric
operator $S : V \supset D(S) = JPD(\Lambda) \rightarrow V$ such that
$$\Lambda f = x(SJPf) + l(f) \quad , \tag{4.4} $$

\nind for every $f \in D(\Lambda)$. Moreover, each such pair $(l,S)$
defines a linear $\Lambda \in \Ls_{CCR}$ in this manner, with 
$D(\Lambda) = D(SJP)$.
\endproclaim

\demo{Proof} Let $\Lambda \in \Ls_{CCR}$. From Theorem 3.1.6 it may be assumed 
that $V \subset D(\Lambda)$ and that $\Lambda$ has the form given in (4.4).
Thus, one sees that
$$\Phi_{\Lambda}(f) = \Phi(f) + \Lambda f \idty = 
\Phi(f) + (x(SJPf) + l(f))\idty = \Phi(f + SJPf) + l(f)\idty \quad , $$

\nind since $R(S) \subset V$. But since $\Lambda \in \Ls_{CCR}$, these 
operators must satisfy the CCR; hence, the operator $\idty + SJP$ must
be symplectic. The resultant equalities
$$\align
\sigma(f,g) &= \sigma(f + SJPf,g + SJPg) \\
  &= \sigma(f,g) + \sigma(f,SJPg) + \sigma(SJPf,g) \\
  &= \sigma(f,g) + \sigma(Pf,SJPg) + \sigma(SJPf,g) \\
  &= \sigma(f,g) + s(JPf,SJPg) - s(SJPf,JPg) \quad , \tag{4.5}
\endalign $$

\nind for all $f,g \in D(\Lambda)$, imply $s(JPf,SJPg) = s(SJPf,JPg)$,
in other words, $S$ is a symmetric operator. The same computation (4.5)
shows that if $S$ is symmetric, then (4.4) defines an element
$\Lambda \in \Ls_{CCR}$.
\hfill\qed\enddemo

\bigpagebreak

\heading V. Canonical Transformations of Finite Degree \endheading

     In this chapter we restrict our attention to the computationally simpler
canonical transformations of arbitrary but finite degree.\footnote{These
are the counterparts in our approach to the polynomial representations of
\cite{30}. Note, however, that due to the boundedness assumptions made in
\cite{30}, which do not need to be made here, we shall be discussing a 
larger class of representations than does \cite{30}. In any case, the 
questions treated below are not addressed in \cite{30} --- at the cost of the
additional technicalities involved in working infinitesimally with 
representations of the CCR, one gains a more detailed computational power
than one apparently can attain when working globally from the outset.} We 
provide necessary and sufficient conditions on 
the mapping $\Lambda$ of finite degree so that the associated 
representation $\pi_{\Lambda}$ of the CCR is unitarily equivalent to a Fock,
a coherent or a quasifree representation. As we show, these results contain
the previously known conditions \cite{33}\cite{9} for the unitary equivalence
of irreducible quasifree representations. The case of unitary equivalence with
a quadratic representation is briefly indicated at the end of the chapter.
We emphasize that when the conditions isolated in this chapter are violated,
one has a representation of the CCR which can describe bosonic systems
with infinitely many degrees of freedom manifesting physics {\it different} 
from that describable by the Fock, coherent, quasifree or quadratic 
representations.

\proclaim{Definition 5.1} Let $\Ls^{(n)}_{CCR}$ denote the set of elements 
$\Lambda$ of $\Ls_{CCR}$ such that

$$\Lambda = \underset{i=0}\to{\overset{n}\to{\sum}} P_i \Lambda \neq
\underset{i=0}\to{\overset{n-1}\to{\sum}} P_i \Lambda \quad . $$

\nind Such elements and the corresponding canonical transformations and
representations will be said to be of degree $n$.
\endproclaim

     Using the equality (3.1.1) given above and the estimate (4.3.5) given
in \cite{22}, the following lemma can be easily proven.

\proclaim{Lemma 5.2} There exist real constants $C_{lm} > 0$ such that

$$\Vert \varphi_l \varphi_m \Vert_2 \leq C_{lm} \Vert \varphi_l \Vert_2
\, \Vert \varphi_m \Vert_2 \quad , $$

\nind for all $\varphi_l \in R(\underset{j \leq l}\to{\sum} P_j)$ and
$\varphi_m \in R(\underset{j \leq m}\to{\sum} P_j)$. Therefore, the 
finite-particle vectors are contained in the domain of the field operators
$\Phi_{\Lambda}(f)$, for all $f \in D(\Lambda)$, whenever 
$\Lambda \in \Ls^{(n)}_{CCR}$.
\endproclaim

     Another straightforward fact we shall need is given in the next lemma.

\proclaim{Lemma 5.3} Let $\Lambda \in \Ls^{(n)}_{CCR}$. Then one has

$$P_m \Phi_{\Lambda}(f) \subset P_m \Phi_{\Lambda}(f) 
\underset{k \leq n+m+1}\to{\sum} P_k \quad , $$

\nind for any $m \in \IN \cup \{ 0 \}$ and $f \in D(\Lambda)$.
\endproclaim

\demo{Proof} For arbitrary $\varphi \in D(\Phi_{\Lambda}(f))$ and
$\psi \in \L2Q$, one sees from Lemma 5.2 that
$$\align
\langle \psi,P_m \Phi_{\Lambda}(f)\varphi\rangle &=
\langle \Phi_{\Lambda}(f) P_m \psi,\varphi\rangle  \\
&= \langle \underset{k \leq \max\{n+m,m+1\}}\to{\sum} P_k
\Phi_{\Lambda}(f) P_m \psi,\varphi\rangle \\
&= \langle \psi,P_m \Phi_{\Lambda}(f) \underset{k \leq n+m+1}\to{\sum} P_k \varphi \rangle \quad .
\endalign $$
\hfill\qed\enddemo

     We give a characterization of the existence of strong graph limits of
sequences of field operators in representations of degree $n$.

\proclaim{Lemma 5.4} Let $\Lambda \in \Ls^{(n)}_{CCR}$ and 
$\{ f_m \}_{m \in \IN} \subset D(\Lambda)$. There exist vectors
$g_m,h_m \in V$ such that $f_m = g_m + Jh_m$, for every $m \in \IN$. The
strong graph limit of the sequence of operators $\{ \Phi_{\Lambda}(f_m)\}$
exists if and only if the sequences $\{ h_m \}$ and
$\{ x(g_m) + \Lambda f_m \}$ converge in their respective Hilbert spaces.
If this strong graph limit exists and the sequence $\{ f_m \}$ converges,
then also the sequence $\{ \Lambda f_m \}$ converges.
\endproclaim

\demo{Proof} Assume that the strong graph limit indicated exists (the other
direction can be proven using the argument of Lemma 3.2.1). Since this limit
is densely defined, for arbitrary $\epsilon > 0$ there exists a sequence
$\{\varphi_m\}_{m \in \IN}$ such that both it and the sequence
$\{\Phi_{\Lambda}(f_m)\varphi_m\}_{m \in \IN}$ converge and such that
$\Vert \varphi_m - \Omega \Vert_2 < \epsilon$, for all $m \in \IN$. Hence,
the sequence $\{P_0 \varphi_m\}_{m \in \IN}$ converges and one has
$\Vert P_0 \varphi_m \Vert > 1 - \epsilon$. One may therefore choose the
sequence $\{\varphi_m\}_{m \in \IN}$ such that, in addition, one has
$P_0 \varphi_m = \Omega$, for all $m \in \IN$. There also exists a real
constant $C$ such that $C \geq C_{kl}$, for $k,l \leq 2n + 2$, where
the constants $C_{kl}$ are those evoked in Lemma 5.2, and thus
$$\Vert \Phi(f) \varphi \Vert_2 \leq C \Vert f \Vert \, \Vert\varphi\Vert_2 \quad , $$
\nind for arbitrary $\varphi$ in the range of the projection 
$\sum_{l \leq 2n+2}P_l$ and $f \in H$. One also has the following
estimate, using previously established notation:
$$\align
\Vert \Phi_{\Lambda}(f_m)\Omega\Vert_2 &= \Vert\Phi(f_m)\Omega +\Lambda f_m \Vert_2 \\
&= \Vert\Phi(Jh_m)\Omega + x(g_m) + \Lambda f_m \Vert_2 \\
&= \Vert ix(h_m) + x(g_m) + \Lambda f_m \Vert_2 \\
&= (\Vert x(h_m) \Vert_2^2+ \Vert x(g_m) + \Lambda f_m \Vert_2^2)^{1/2} \\
&\overset{(3.1.1)}\to{=} 
(\frac{1}{2}\Vert h_m \Vert_2^2+ \Vert x(g_m) + \Lambda f_m \Vert_2^2)^{1/2} \\
&\geq \frac{1}{2} \max \{ \Vert h_m \Vert,\Vert x(g_m) + \Lambda f_m \Vert_2\} \\
&\geq \frac{1}{4}(\Vert h_m \Vert + \Vert x(g_m) + \Lambda f_m \Vert_2 )
\quad . \tag{5.1} 
\endalign $$
\nind Of course, also the sequence 
$\{ \sum_{j=0}^n P_j \Phi_{\Lambda}(f_m)\varphi_m \}_{m \in \IN}$ converges.
Now consider the estimate (obtained using Lemma 5.3)
$$\align
\quad &\Vert \underset{j=0}\to{\overset{n+1}\to{\sum}} P_j \Phi_{\Lambda}(f_m)
\varphi_m - \Phi_{\Lambda}(f_m)\Omega \Vert_2 
  = \Vert \underset{j=0}\to{\overset{n+1}\to{\sum}} P_j \Phi_{\Lambda}(f_m)
\underset{l=1}\to{\overset{2n+2}\to{\sum}} P_l\varphi_m\Vert_2 \\ 
&\leq \Vert \underset{j=0}\to{\overset{n+1}\to{\sum}} P_j \Phi(Jh_m)
\underset{l=1}\to{\overset{2n+2}\to{\sum}} P_l\varphi_m\Vert_2 +
\Vert \underset{j=0}\to{\overset{n+1}\to{\sum}} P_j (x(g_m)+\Lambda f_m)
\underset{l=1}\to{\overset{2n+2}\to{\sum}} P_l\varphi_m\Vert_2 \\ 
&\leq \Vert \Phi(Jh_m)
\underset{l=1}\to{\overset{2n+2}\to{\sum}} P_l\varphi_m\Vert_2 +
\Vert (x(g_m)+\Lambda f_m)
\underset{l=1}\to{\overset{2n+2}\to{\sum}} P_l\varphi_m\Vert_2 \\ 
&\leq C(\Vert h_m \Vert + \Vert x(g_m)+\Lambda f_m \Vert_2) \,
\Vert \underset{l=1}\to{\overset{2n+2}\to{\sum}} P_l\varphi_m\Vert_2 \\ 
&\leq C(\Vert h_m \Vert + \Vert x(g_m)+\Lambda f_m \Vert_2)\,\epsilon \\
&\overset{(5.1)}\to{\leq} \Vert \Phi_{\Lambda}(f_m)\Omega\Vert_2 \, 4 C \epsilon \quad . 
\endalign $$
\nind From this estimate, one sees that the boundedness of the sequence\newline
$\{ \underset{j=0}\to{\overset{n+1}\to{\sum}} P_j \Phi_{\Lambda}(f_m)
\varphi_m \}_{m \in \IN}$ entails the boundedness of the sequence
$\{ \Vert \Phi_{\Lambda}(f_m)\Omega \Vert_2\}_{m \in \IN}$. Thus, for any
$\delta > 0$, there exist convergent sequences $\{\varphi_m\}_{m \in \IN}$ and
\newline $\{ \underset{j=0}\to{\overset{n+1}\to{\sum}} P_j \Phi_{\Lambda}(f_m)
\varphi_m \}_{m \in \IN}$ such that

$$\Vert\underset{j=0}\to{\overset{n+1}\to{\sum}} P_j \Phi_{\Lambda}(f_m)
\varphi_m - \Phi_{\Lambda}(f_m)\Omega \Vert_2 < \delta \quad , $$

\nind for all $m \in \IN$. This implies the convergence of the sequence
$\{ \Phi_{\Lambda}(f_m)\Omega\}$. It may therefore be concluded that the
sequences $\{ h_m \}_{m \in \IN}$ and $\{ x(g_m) + \Lambda f_m \}_{m \in \IN}$
converge (use (5.1) with $f_m$ replaced by $f_{m_1} - f_{m_2}$).
\hfill\qed\enddemo

     It is easy to see that the following lemma is true. We simply record it
here for later reference.

\proclaim{Lemma 5.5} Let $\pi$ be a representation of a $C^*$-algebra
$\As$, $\{ A_n \}$ a sequence of elements of $\As$, and $k$ a cardinal 
number. The strong graph limit of the operator sequence
$\{ \pi(A_n) \}$ exists and is densely defined if and only if
the strong graph limit of the operator sequence $\{ k \pi(A_n) \}$ exists and 
is densely defined, where $k\pi$ is the direct sum of $k$ copies of $\pi$.
\endproclaim

     We can finally prove our characterization of the quasi-equivalence of a
representation $\pi_{\Lambda}$ of degree $n$ with the original Fock 
representation $\pi_J$. This generalizes Theorem 4.1 in \cite{22}, which was
restricted to the case $n=2$. Of course, if $\pi_{\Lambda}$ is irreducible
(see Theorem 3.2.2), then quasi-equivalence implies unitary equivalence.    

\proclaim{Theorem 5.6} Let $\Lambda \in \Ls^{(n)}_{CCR}$. The representation
$\pi_{\Lambda}$ is quasi-equivalent to the restriction of the Fock
representation $\pi_J$ to $\As(D(\Lambda))$ if and only if the operator
$\overline{\Lambda}$ is Hilbert-Schmidt.
\endproclaim

\demo{Proof} Assume that the representations $\pi_{\Lambda}$ and
$\pi_J \mid_{\As(D(\Lambda))}$ are quasi-equivalent. Let $f \in H$ and 
$\{ f_n \}_{n \in \IN} \subset D(\Lambda)$ be arbitrary with 
$\{ f_n \}_{n \in \IN}$ converging to $f$. The convergence of the sequence
$\{ \Phi(f_n)\varphi \}_{n \in \IN}$ to $\Phi(f)\varphi$ for finite-particle
vectors $\varphi$ entails that $\Phi(f)$ is the strong graph limit of 
$\{ \Phi(f_n) \}_{n \in \IN}$ (see Theorem VIII.27 in \cite{24}).
As quasi-equivalence is the same as unitary equivalence up to multiplicity
(see {\it e.g.} Theorem 2.4.26 in \cite{5}), Lemma 5.5 implies that the
strong graph limit of $\{ \Phi_{\Lambda}(f_n)\}_{n \in \IN}$ exists, so that,
by Lemma 5.4, the sequence $\{ \Lambda f_n \}_{n \in \IN}$ converges. This
entails that $\Lambda$ is bounded, and thus, by Lemma 3.2.1, one may assume
that $D(\Lambda) = H$. According to Theorem 5.2.14 in \cite{6} there 
must exist a dense subset $\Ks$ of $\L2Q$ such that
$$\underset{k=1}\to{\overset{\infty}\to{\sum}} \Vert (\Phi_{\Lambda}(e_k) +
i\Phi_{\Lambda}(Je_k))\varphi\Vert_2^2 < \infty \quad , \tag{5.2} $$
\nind for all $\varphi \in \Ks$, since a densely defined number operator
exists. Thus, one must have
$$\align
&\underset{k=1}\to{\overset{\infty}\to{\sum}} \Vert 
\underset{l=0}\to{\overset{n}\to{\sum}} P_l (\Phi_{\Lambda}(e_k) +
i\Phi_{\Lambda}(Je_k))\varphi\Vert_2^2 \\
&=
\underset{k=1}\to{\overset{\infty}\to{\sum}} \Vert 
\underset{l=0}\to{\overset{n}\to{\sum}} P_l (\Phi_{\Lambda}(e_k) +
i\Phi_{\Lambda}(Je_k))
\underset{m=0}\to{\overset{2n+1}\to{\sum}} P_m \varphi\Vert_2^2 \\
&< \infty \quad , 
\endalign $$
\nind and, since the finite-particle vectors are in the domain of the
number operator of the Fock representation $\pi_J$, 
$$\underset{k=1}\to{\overset{\infty}\to{\sum}} \Vert 
\underset{l=0}\to{\overset{n}\to{\sum}} P_l (\Phi_{\Lambda}(e_k) +
i\Phi_{\Lambda}(Je_k) - \sqrt{2} a(e_k))
\underset{m=0}\to{\overset{2n+1}\to{\sum}} P_m \varphi\Vert_2^2 
< \infty \quad . \tag{5.3} $$

\nind For each $\epsilon > 0$ there exists a vector 
$\varphi_{\epsilon} \in \Ks$ such that 
$\Vert \varphi_{\epsilon} - \Omega\Vert_2 < \epsilon$; and set

$$C \equiv \underset{r,s \leq 2n+1}\to{\max}C_{rs}$$

\nind (see Lemma 5.2). Then, since 
$\Vert (\Lambda e_k + i\Lambda Je_k) P_0 \varphi_{\epsilon}\Vert_2 =
\Vert (\Lambda e_k + i\Lambda Je_k)\Vert_2 \Vert P_0 \varphi_{\epsilon}\Vert_2$
($P_0 \varphi_{\epsilon}$ is just a constant function), one has the estimate
$$\align
\Vert 
\underset{l=0}\to{\overset{n}\to{\sum}} &P_l (\Phi_{\Lambda}(e_k) +
i\Phi_{\Lambda}(Je_k) - \sqrt{2} a(e_k))
\underset{m=0}\to{\overset{2n+1}\to{\sum}} P_m \varphi_{\epsilon}\Vert_2 \\
&= \Vert \underset{l=0}\to{\overset{n}\to{\sum}} P_l (\Lambda e_k +
i\Lambda Je_k) 
\underset{m=0}\to{\overset{2n+1}\to{\sum}} P_m \varphi_{\epsilon}\Vert_2 \\
&\geq
\Vert \underset{l=0}\to{\overset{n}\to{\sum}} P_l (\Lambda e_k +
i\Lambda Je_k) P_0 \varphi_{\epsilon}\Vert_2 -
\underset{m=1}\to{\overset{2n+1}\to{\sum}}
\Vert \underset{l=0}\to{\overset{n}\to{\sum}} P_l (\Lambda e_k +
i\Lambda Je_k) P_m \varphi_{\epsilon}\Vert_2 \\
&\geq
\Vert (\Lambda e_k + i\Lambda Je_k) P_0 \varphi_{\epsilon}\Vert_2 -
\underset{m=1}\to{\overset{2n+1}\to{\sum}}
\Vert (\Lambda e_k + i\Lambda Je_k) P_m \varphi_{\epsilon}\Vert_2 \\
&\geq 
\Vert \Lambda e_k + i\Lambda Je_k \Vert_2 \Vert P_0 \varphi_{\epsilon}\Vert_2 
- \underset{m=1}\to{\overset{2n+1}\to{\sum}} C
\Vert \Lambda e_k + i\Lambda Je_k \Vert_2 \Vert P_m \varphi_{\epsilon}\Vert_2 \\
&\geq (1-\epsilon) \Vert \Lambda e_k + i\Lambda Je_k \Vert_2 -
(2n+1)C\epsilon \Vert \Lambda e_k + i\Lambda Je_k \Vert_2 \\
&= (1-\epsilon - (2n+1)C\epsilon)\Vert \Lambda e_k + i\Lambda Je_k \Vert_2
\quad . \tag{5.4}
\endalign $$

\nind Choosing $\epsilon > 0$ such that
$1-\epsilon - (2n+1)C\epsilon \geq \frac{1}{2}$, one sees that

$$\underset{k}\to{\sum} \Vert \Lambda e_k + i\Lambda Je_k \Vert_2^2 < \infty
\quad , $$

\nind so that $\Lambda$ is a Hilbert-Schmidt operator ($\Lambda e_k$ and
$\Lambda Je_k$ are a.e. real-valued):

$$\underset{k}\to{\sum} (\Vert\Lambda e_k \Vert_2^2 + \Vert\Lambda Je_k \Vert_2^2) 
= \underset{k}\to{\sum} \Vert \Lambda e_k + i\Lambda Je_k \Vert_2^2 \quad . $$

     The asserted sufficiency of the condition will now be proven. Without loss
of generality, one may assume that $\Lambda = \overline{\Lambda}$ and 
$P_0 \Lambda = 0$. There exist real constants $\lambda_{kk_1 \cdots k_m}$ 
symmetric in the indices, such that (using (3.1.1))

$$\Lambda Je_k = \underset{m,k_1,\ldots,k_m}\to{\sum}\lambda_{kk_1 \cdots k_m}
:x_{k_1} \cdots x_{k_m}:$$
and
$$\underset{k}\to{\sum}\underset{m,k_1,\ldots,k_m}\to{\sum}\lambda^2_{kk_1 \cdots k_m}\frac{m!}{2^m} =
\underset{k}\to{\sum} \Vert\Lambda Je_k\Vert_2^2 < \infty \quad . $$

\nind Setting 

$$G = \underset{m,k_1,\ldots,k_m}\to{\sum}\frac{\lambda_{k_1 \cdots k_m}}{m}
:x_{k_1} \cdots x_{k_m}: \quad \in \L2Q \quad , $$

\nind one sees that

$$e^{-iG}\Phi(f)e^{iG} = \Phi(f) \quad , $$

\nind for all $f \in V$, as well as

$$e^{-iG}\Phi(Je_k)e^{iG} = \Phi_{\Lambda}(Je_k) \quad , $$

\nind for all $k \in \IN$, by Proposition 3.1.3. This then demonstrates that

$$e^{-iG}\Phi(g)e^{iG} = \Phi_{\Lambda}(g) \quad , $$

\nind for all $g$ in the linear span of $\{ Je_k \mid k \in \IN \}$. Lemma
3.2.1 completes the proof.
\hfill\qed\enddemo

     And next we give a characterization of the quasi-equivalence of a 
representation of degree $n$ with a quasifree representation. 

\proclaim{Theorem 5.7} Let $\Lambda \in \Ls^{(n)}_{CCR}$ and $\pi$ be a
GNS-representation of a pure quasifree state on $\As(D(\Lambda))$. There
exists a symplectic operator $K : D(\Lambda) \rightarrow H$ such that
$\Phi_{P_1 \Lambda}(f) = \Phi(Kf)$, for all $f \in D(\Lambda)$. The
representations $\pi_{\Lambda}$ and $\pi$ are quasi-equivalent if and only if
the following conditions are fulfilled: 

   (i) $\pi_{(P_0 + P_1)\Lambda}$ and $\pi$ are quasi-equivalent.

   (ii) The closure of the restriction of 
$(\Lambda - P_0 \Lambda - P_1 \Lambda)_{max}$ to the range of $K$ is
Hilbert-Schmidt.
\endproclaim

\demo{Proof} By Lemma 4.1 and Corollary 4.4, it may be assumed that $\pi$ is 
of the form

$$\pi(W(f)) = e^{i(\Phi(Tf)+l(f))} \quad , \quad f \in D(\Lambda) \quad , $$

\nind for a symplectic operator $T : D(\Lambda) \rightarrow H$ with
dense range and a linear form $l : D(\Lambda) \rightarrow \RR$. After a coherent
transformation, one may assume that $l$ is the zero mapping. Let the
representations $\pi$ and $\pi_{\Lambda}$ be quasi-equivalent. Then the
restriction of $\pi_J$ to $\As(R(T))$ and the representation $\pi'$, defined
by

$$\pi'(W(f)) = e^{i\Phi_{\Lambda}(T^{-1}f)} \quad , \quad f \in R(T) \quad , $$

\nind are quasi-equivalent. The linear part of the field operators associated
with $\pi'$ is, up to constants, equal to $\Phi_{P_1 \Lambda}(T^{-1}f)$.
There exists a symplectic transformation $S : R(T) \rightarrow H$ such that
$\Phi(Sf) = \Phi_{P_1 \Lambda}(T^{-1}f)$. As in the proof of Theorem 5.6,
one may extend $\pi'$ to a representation of $\As(H)$ which is 
quasi-equivalent to $\pi_J$. This extension will also be denoted by $\pi'$.
The operator $\Lambda \circ T^{-1}$ is bounded; consequently $S$ is bounded,
as well. It may be assumed that $S = \overline{S}$. \par
     Proceeding as in the proof of Theorem 5.6, one needs to consider the
linear part of the annihilation operators associated with $\pi'$ separately.
Furthermore, the operators $a(e_k)$ in equation (5.3) must be replaced by
somewhat different terms, discussed below. Let $\Phi'(f)$, $f \in H$, denote 
the field operators associated with $\pi'$. The linear part of the associated 
annihilation operators $\frac{1}{\sqrt{2}}(\Phi'(e_k)+i\Phi'(Je_k))$ is 
given by
$$\align
\frac{1}{\sqrt{2}}(\Phi(Se_k) + i\Phi(SJe_k)) &= 
\frac{1}{2}(a(Se_k) + a^*(Se_k) + ia(SJe_k) + ia^*(SJe_k)) \\
&= \frac{1}{2}(a((S-JSJ)e_k) + a^*((S + JSJ)e_k)) \quad . \tag{5.5}
\endalign $$

\nind The operators $S - JSJ$ and $J$ commute, so that, if 
$S-JSJ = UA$ is the polar decomposition, then $U$ and $J$ commute. Moreover,
since $R(A) = \Hs$, one has $U^* U = \idty$. Therefore, (5.3) is still 
fulfilled if $a(e_k)$ is replaced by $a(Ue_k)$. \par
     The bound 
$$\Vert (S-JSJ-2U)f\Vert \leq \Vert (S+JSJ)f \Vert \quad , \tag{5.6} $$

\nind will be proven for every $f \in H$. Note that for an arbitrary unit
vector $f \in H$, one has
$$\align
\Vert Af \Vert^2 &= \Vert (S-JSJ)f \Vert^2 \\
  &= \Vert Sf \Vert^2 + \Vert JSJf \Vert^2 - 2\langle Sf,JSJf\rangle \\
  &= \Vert Sf \Vert^2 + \Vert SJf \Vert^2 + 2\sigma(Sf,SJf) \\
  &= \Vert Sf \Vert^2 + \Vert SJf \Vert^2 + 2\sigma(f,Jf) \\
  &= \Vert Sf \Vert^2 + \Vert SJf \Vert^2 + 2  \quad . 
\endalign $$

\nind Similarly, it follows that

$$\Vert (S+JSJ)f \Vert^2 = \Vert Sf \Vert^2 + \Vert SJf \Vert^2 - 2 \quad . $$

\nind From the bound

$$1 = \vert\sigma(SJf,Sf)\vert \leq \Vert SJf \Vert \, \Vert Sf \Vert $$

\nind it follows that

$$\Vert Af \Vert^2 \geq \Vert Sf \Vert^2 + \frac{1}{\Vert Sf \Vert^2} + 2 
\geq 4 \quad , $$

\nind and $A \geq 2\idty$. This proves (5.6), as $A \geq 2\idty$ and
$U^* U = \idty$ imply
$$\align
\Vert (S-JSJ-2U)f\Vert^2 &= \Vert (A - 2\idty)f \Vert^2 \\
  &= \langle f,(A^2 - 4A + 4\idty)f\rangle \\
  &\leq \langle f,(A^2 - 8\idty + 4\idty)f\rangle \\
  &= \Vert (S - JSJ)f \Vert^2 - 4 \\
  &= \Vert (S + JSJ)f \Vert^2 \quad .
\endalign $$

     One can then proceed as in the proof of Theorem 5.6 to find suitable
Hilbert-Schmidt conditions - one must simply modify (5.5) by the term
$-a(Ue_k)$, resp., replace $a(e_k)$ in (5.3) by $a(Ue_k)$. In the counterpart
to (5.4) one finds the additional terms
$$\align
\Vert a^*((S+JSJ)e_k)P_0 \varphi_{\epsilon}\Vert_2 &\overset{(3.1.1)}\to{=} 
\Vert(S+JSJ)e_k\Vert\,\Vert P_0 \varphi_{\epsilon}\Vert_2 \quad , \\
\Vert a^*((S+JSJ)e_k)\underset{l=1}\to{\overset{2n+1}\to{\sum}}P_l 
\varphi_{\epsilon}\Vert_2 &\leq C\epsilon\Vert(S+JSJ)e_k \Vert \quad , 
\endalign $$
and
$$\align
\Vert a((S-JSJ-2U)&e_k)\underset{l=1}\to{\overset{2n+1}\to{\sum}}P_l 
\varphi_{\epsilon}\Vert_2 \\
&\leq C\Vert (S-JSJ-2U)e_k \Vert \,\Vert \underset{l=1}\to{\overset{2n+1}\to{\sum}}P_l 
\varphi_{\epsilon}\Vert_2 \\
&\overset{(5.6)}\to{\leq} \quad C \Vert(S+JSJ)e_k \Vert \epsilon \quad .
\endalign $$

\nind It may therefore be concluded that $S + JSJ$ is a Hilbert-Schmidt
operator. Furthermore, the constant part of the field operator $\Phi'(f)$ is
a bounded linear form, so by Lemma 5.8 it may be assumed without loss of
generality that it is trivial. Thus, the representations $\pi$ and 
$\pi_{P_0 \Lambda +P_1 \Lambda}$ are quasi-equivalent. The closure
of the operator $(\Lambda - P_0 \Lambda - P_1 \Lambda) \circ T^{-1}$ is also
Hilbert-Schmidt. That the mapping $S^{-1} : R(S) \rightarrow D(S) = H$ is bounded
is implied by the bound

$$\Vert Sf \Vert \,\Vert S \Vert \,\Vert f \Vert \geq \vert\sigma(Sf,SJf)\vert
= \Vert f \Vert^2 \quad , $$

\nind for every $f \in H$, as well as

$$\Vert Sf \Vert \geq \frac{1}{\Vert S \Vert}\Vert f \Vert \quad . $$

\nind Hence, recalling $S = \overline{KT^{-1}}$, the operator

$$\align
\overline{(\Lambda - P_0 \Lambda - P_1 \Lambda) \circ T^{-1}} \circ S^{-1}
&= \overline{(\Lambda - P_0 \Lambda - P_1 \Lambda) \circ T^{-1} \circ T \circ K^{-1}} \\
&= \overline{(\Lambda - P_0 \Lambda - P_1 \Lambda)_{max} \circ P \circ K^{-1}} \\
&= \overline{(\Lambda - P_0 \Lambda - P_1 \Lambda)_{max} \mid R(K)}
\endalign $$

\nind is Hilbert-Schmidt. \par
     To prove that the stated Hilbert-Schmidt conditions imply the desired
quasi-equivalence, it is sufficient to show that the representation admits
a densely defined number operator, in other words, by Theorem 5.2.14 in
\cite{6} it suffices to show that for
any finite-particle vector $\varphi = \sum_{l=0}^m P_l \varphi$, and any
orthonormal basis $\{ f_k,Jf_k \mid k \in \IN \}$ of $H$, one has

$$\align
\underset{k}\to{\sum} &\Vert (\Phi'(f_k) + i\Phi'(Jf_k))\varphi - 
\sqrt{2}a(Uf_k)\varphi\Vert_2^2 \\
&\leq D^{(m)}(\Vert S+JSJ \Vert_{HS}^2 + 
\Vert \overline{(\Lambda - P_0 \Lambda - P_1 \Lambda)_{max} \mid R(K)} \Vert_{HS}^2) \,\Vert \varphi \Vert_2^2 \\
&< \infty  \quad ,
\endalign $$

\nind for a constant $D^{(m)} \in \RR$ which does not depend on the choice
of the basis (the Hilbert-Schmidt norm of an operator is denoted by
$\Vert\cdot\Vert_{HS}$). This can be done by using arguments already employed
above (see Lemma 5.2 and equations (5.5) and (5.6)). It follows then that

$$\underset{k}\to{\sum} \Vert(\Phi'(f_k)+i\Phi'(Jf_k))\varphi\Vert^2$$

\nind is bounded by a finite real number, which only depends on $m$,
$\Vert\varphi\Vert_2$, $\Vert S+JSJ \Vert_{HS}^2$, and
$\Vert \overline{(\Lambda - P_0 \Lambda - P_1 \Lambda)_{max} \mid R(K)} \Vert_{HS}^2$. 
\hfill\qed\enddemo

     Taken together, Theorems 5.6 and 5.7 give conditions so that
our higher order representations are {\it inequivalent} to the well-studied
class of representations associated with inhomogeneous linear canonical
transformations, and hence so that they can describe bosonic systems with
{\it different} physics. \par

\proclaim{Lemma 5.8} Let $\pi$ be a GNS-representation of a quasifree
state $\omega$ on $\As(H_0)$, where $H_0$ is a dense subspace of $\Hs$.
There exists a linear form $k : H_0 \rightarrow \RR$ and a scalar product
$s'$ on $H_0$ such that 

$$\omega(W(f)) = e^{ik(f) - \frac{s'(f,f)}{4}} \quad , $$

\nind for all $f \in H_0$ \cite{18}. If $l : H_0 \rightarrow \RR$ is a linear
form, then the representation $\pi_l$ of $\As(H_0)$ defined by 
$\pi_l(W(f)) = e^{il(f)}\pi(W(f))$ is quasi-equivalent to $\pi$ if
and only if $l$ is bounded with respect to $s'$.
\endproclaim

\demo{Proof} By Corollary 4.4, it may be assumed that the linear form $k$ is
trivial and that $\pi(W(f)) = e^{i\Phi(Tf)}$, $f \in H_0$, for a symplectic
mapping $T : H_0 \rightarrow H$. Suppose the indicated representations are
quasi-equivalent. The convergence $f_n \overset{s'}\to{\rightarrow} 0$ implies
the convergence $Tf_n \overset{s}\to{\rightarrow} 0$, so that the strong graph
limit of $\{\Phi(Tf_n)\}$ is $0$. The assumed quasi-equivalence implies also
that $0$ is the strong graph limit of the sequence $\{\Phi(Tf_n)+l(f_n)\}$.
Appealing to Lemma 5.4, one concludes that also $\{l(f_n)\}$ converges to
$0$. \par
     Assume now that the linear form $l$ is bounded with respect to $s'$,
so that $l(f) = s'(g,f)$ for some element $g$ in the completion of $H_0$ with
respect to $s'$. There exists a sequence $\{ g_n \} \subset H_0$ converging
to $g$ with respect to $s'$; hence the sequence $\{ Tg_n \}$ converges to
some element $h \in H$ with respect to $s$. Then, for $f \in H_0$, one sees
$$\align
e^{i\Phi(Jh)}\Phi(Tf)e^{-i\Phi(Jh)} &= \Phi(Tf) - \sigma(Jh,Tf)\idty \\
&=  \Phi(Tf) + s(h,Tf)\idty \\
&=  \Phi(Tf) + \underset{n}\to{\lim} \, s(Tg_n,Tf) \idty\\
&=  \Phi(Tf) + s'(g,f) \idty\\
&=  \Phi(Tf) + l(f)\idty \quad . 
\endalign $$
\hfill\qed\enddemo

     We provide an extension to Proposition 3.1.7 for irreducible representations
of finite degree.

\proclaim{Proposition 5.9} Let $\Lambda \in \Ls^q_{CCR} \cap \Ls^{(n)}_{CCR}$
determine an irreducible representation $\pi_{\Lambda}$ of $\As(D(\Lambda))$.
If $l : H_0 \rightarrow \RR$ is a linear form such that the representation $\pi_l$ 
of $\As(H_0)$ defined by $\pi_l(W(f)) = e^{il(f)}\pi_{\Lambda}(W(f))$ is 
unitarily equivalent to $\pi_{\Lambda}$, then $l$ has the form
$l(f) = \sigma(g,f)$, for all $f \in D(\Lambda)$, for some 
$g \in D(\Lambda_{max})$. Therefore, the set $D(\Lambda_{max})$ is the 
maximal extension of the test function space $D(\Lambda)$.
\endproclaim

\demo{Proof} Let $U : \L2Q \rightarrow \L2Q$ be unitary such that 

$$U^* \Phi_{\Lambda}(f)U = \Phi_{\Lambda}(f) + l(f)\idty \quad , $$

\nind for all $f \in D(\Lambda)$. Then the operator
$\pi_{\Lambda_{max}}(W(g))U^*\pi_{\Lambda_{max}}(W(-g))U$ commutes with the
elements of $\pi_{\Lambda}(\As(D(\Lambda)))$, for all $g \in D(\Lambda_{max})$.
By the assumed irreducibility, it must be a multiple of the identity. Thus,
for each $g \in D(\Lambda_{max})$ one has

$$U^*\pi_{\Lambda_{max}}(W(g))U = e^{ic_g}\pi_{\Lambda_{max}}(W(g)) \quad , $$

\nind for suitable $c_g \in \RR$. \par
     It shall next be shown that $e^{ic_g} = e^{il'(g)}$, for a suitable linear
form $l' : D(\Lambda_{max}) \rightarrow \RR$. But the equality
$e^{ic_{t_1 g}}e^{ic_{t_2 g}} = e^{ic_{(t_1 + t_2)g}}$ implies
$e^{ic_{t g}} = e^{itk}$, for all $t \in \RR$ and a suitable $k \in \RR$. 
Thus, it has been shown that

$$U^* \Phi_{\Lambda_{max}}(tg)U = \Phi_{\Lambda_{max}}(tg) + tk \idty\quad , $$

\nind in other words,

$$U^* \pi_{\Lambda_{max}}(W(f))U = e^{il'(f)}\pi_{\Lambda_{max}}(W(f)) \quad , $$

\nind for all $f \in D(\Lambda_{max})$, with 
$l' : D(\Lambda_{max}) \rightarrow \RR$ linear. Hence, there exists an extension
$l' : D(\Lambda_{max}) \rightarrow \RR$ of $l$ such that the representation
$\pi_{l'}$ of $\As(D(\Lambda_{max}))$ defined by 
$\pi_{l'}(W(f)) = e^{il'(f)}\pi_{\Lambda_{max}}(W(f))$, for 
$f \in D(\Lambda_{max})$, is unitarily equivalent to $\pi_{\Lambda_{max}}$.
Therefore, it may be assumed that $l = l'$ and $\Lambda = \Lambda_{max}$. \par
     Let $\{ e_k,Je_k \mid k \in \IN \} \subset D(\Lambda)$ be an orthonormal
symplectic basis in $H$. For each $f \in D((P_m \Lambda)_{max})$, there 
exist real constants $\lambda_{k_1 \cdots k_m}(f)$, which are totally 
symmetric in the indices, such that

$$(P_m \Lambda)_{max} f = \underset{k_1,\ldots,k_m}\to{\sum}
\lambda_{k_1 \cdots k_m}(f) :x_{k_1} \cdots x_{k_m}: \quad . $$

\nind If $\{ f_n \}$ is a convergent sequence in $V$, then the strong graph
limit of $\{ \Phi(f_n)\}$ exists, so that also the strong graph limit of 
$\{ \Phi(f_n) + l(f_n) \}$ exists. Therefore, the restriction of $l$ to
$V$ is bounded. Set $l_k \equiv l(e_k)$, $h \equiv \sum_k l_k Je_k \, \in H$
and 

$$A_l(f) \equiv \Phi_{\Lambda}(f) - \underset{m}\to{\sum}\underset{k_1,\ldots,k_m \leq l}\to{\sum}
\lambda_{k_1 \cdots k_m}(f) :x_{k_1} \cdots x_{k_m}:\idty \quad , $$

\nind with $D(A_l(f)) = D(\Phi_{\Lambda}(f)) \cap (\cap_{k_1,\ldots,k_m \leq l}
D(x_{k_1} \cdots x_{k_m}\idty))$. Then the operator $A_l(f)$ is symmetric and
the sequence $\{ A_l(f)\varphi\}_{l \in \IN}$ converges to $\Phi(f)\varphi$
for every $\varphi \in \Gs$. Since $\Gs$ is a core for $\Phi(f)$ (Lemma 3.1.2),
it follows from Theorem VIII.27 of \cite{24} that $\Phi(f)$ is the
strong graph limit of $\{ A_l(f)\}_{l \in \IN}$. \par
     With $U^* x_k U = x_k + l_k$ and $p_l(f)$ suitable polynomials of
degree $n-2$, one sees therefore that the strong graph limit of
$$\align
U^*&A_l(f)U \\
&= U^* \Phi_{\Lambda}(f)U - \underset{m=2}\to{\overset{n}\to{\sum}} \quad
\underset{k_1,\ldots,k_m \leq l}\to{\sum}
\lambda_{k_1 \cdots k_m}(f) U^* :x_{k_1} \cdots x_{k_m}: U \\
&= \Phi_{\Lambda}(f) + (l(f) + p_l(f)  - 
\underset{k_1,\ldots,k_{n-1} \leq l}\to{\sum}
\lambda_{k_1 \cdots k_{n-1}}(f) :x_{k_1} \cdots x_{k_{n-1}}:)\idty \\
&- \underset{k_1,\ldots,k_n \leq l}\to{\sum}
(\lambda_{k_1 \cdots k_{n}}(f) :x_{k_1} \cdots x_{k_{n}}: +
\lambda_{k_1 \cdots k_{n}}(f) \, n \, l_{k_n} :x_{k_1} \cdots x_{k_{n-1}}:)\idty 
\tag{5.7} \endalign $$
\nind is $U^* \Phi(f) U$. According to the proof of Lemma 5.4, the existence of
the indicated strong graph limit entails the convergence of 
$$\align
p_l(f) \,  &- 
\underset{k_1,\ldots,k_{n-1} \leq l}\to{\sum}
\lambda_{k_1 \cdots k_{n-1}}(f) :x_{k_1} \cdots x_{k_{n-1}}: \\
&- \underset{k_1,\ldots,k_n \leq l}\to{\sum}
(\lambda_{k_1 \cdots k_{n}}(f) :x_{k_1} \cdots x_{k_{n}}: +
\lambda_{k_1 \cdots k_{n}}(f) \, n \, l_{k_n} :x_{k_1} \cdots x_{k_{n-1}}:) \quad . 
\endalign $$
\nind Following the argument of Lemma 3.1, it can also be shown that the 
strong graph limit of $\{ U^* A_l(f) U\}_{l\in\IN}$ is of the form
$\overline{\Phi(f) + F\idty}$, for a suitable 
$F \in R(\sum_{k=0}^{n-1}\, P_k) \subset \L2Q$. Hence, there exists a
$\Lambda' \in \Ls_{CCR}^{(n-1)}$ such that 
$$U^* \Phi(f)U = \Phi_{\Lambda'}(f) \quad , $$
\nind for all $f \in D(\Lambda) = D(\Lambda')$. \par
     According to Theorem 5.6, $\Lambda'$ is a Hilbert-Schmidt operator.
Hence one sees that
$$\align
\infty &> \underset{k}\to{\sum} \Vert P_{n-1}\Lambda'(Je_k) \Vert_2^2 \\
&= \underset{k}\to{\sum}\Vert \underset{k_1,\ldots,k_n}\to{\sum}
\lambda_{k_1 \cdots k_n}(Je_k) l_{k_n}n :x_{k_1} \cdots x_{k_{n-1}}: \Vert_2^2 \\
&\overset{(3.1.1)}\to{=} \underset{k}\to{\sum} \underset{k_1,\ldots,k_{n-1}}\to{\sum} ( \underset{k_n}\to{\sum}
\lambda_{k_1 \cdots k_n}(Je_k) l_{k_n}n)^2 \, \frac{(n-1)!}{2^{n-1}} \\
&= n^2\frac{(n-1)!}{2^{n-1}} \underset{k}\to{\sum} \underset{k_1,\ldots,k_{n-1}}\to{\sum} ( \underset{k_n}\to{\sum}
\lambda_{k_1 \cdots k_{n}}(Je_k) l_{k_n})^2 \\
&\overset{(\text{Lemma} \, 3.1.1)}\to{=} n \frac{n!}{2^{n-1}} \underset{k}\to{\sum} 
\underset{k_1,\ldots,k_{n-1}}\to{\sum} ( \underset{k_n}\to{\sum}\frac{2^n}{n!}
\langle\Lambda Je_k , :x_{k_1} \cdots x_{k_{n-1}}x_{k_n}:\rangle l_{k_n})^2 \\
&= 2n \underset{k,k_1,\ldots,k_{n-1}}\to{\sum} \frac{2^n}{n!}
\langle\Lambda Je_k , :x_{k_1} \cdots x_{k_{n-1}} x(-Jh):\rangle^2 \quad , 
\endalign $$

\nind which entails $h \in D((P_n \Lambda)_{max})$ (see the proof of Proposition 
3.1.7). \par
     There exists a $G \in \L2Q \cap R(P_{n+1})$ such that
$\Phi_{(P_n \Lambda)_{max}}(h) = e^{-iG} \Phi(h) e^{iG}$. By replacing
$U$ by $e^{iG}Ue^{-iG}$ and $\Phi_{\Lambda}(f)$ by 
$e^{iG}\Phi_{\Lambda}(f)e^{-iG}$, which is the same as replacing $\Lambda$
by another element of $\Ls_{CCR}$ differing from $\Lambda$ only in the
component of degree $n$, it may be assumed that 
$(P_n \Lambda)_{max}h = 0$ and hence that the range of the mapping
$(P_n \Lambda)_{max}$ lies in the subspace of $\L2Q$ generated by

$$\{ : x(f_1) \cdots x(f_n): \mid \sigma(h,f_1) = \ldots = \sigma(h,f_n) = 0,
\, f_1, \ldots, f_n \in V \} $$

\nind (use (3.1.3) and the fact that $a(h)(P_n \Lambda)_{max}f = 0$, for
$f \in D((P_n \Lambda)_{max})$). The operator $e^{i\Phi(h)}$ commutes with
these elements. From the definition of $h$, the adjoint actions of 
$e^{i\Phi(h)}$ and $U$ on the field operators $\Phi_{\Lambda}(f) =\Phi(f)$,
$f \in V$, are identical, inducing the same coherent canonical transformation.
Hence, $e^{-i\Phi(h)}U$ commutes with $e^{ix(f)}$, for all $f \in V$, and
thereby may be identified with the multiplication operator corresponding to
some suitable element of $L^{\infty}(Q,d\mu)$, as in the proof of Theorem 
3.2.2. Thus, one has

$$U^*((P_n \Lambda)_{max} f)U = (P_n \Lambda)_{max} f \quad , $$

\nind for all $f \in D((P_n \Lambda)_{max})$, and it therefore follows that

$$U^* \Phi_{(\Lambda - P_n \Lambda)_{max}}(f)U = 
\Phi_{(\Lambda - P_n \Lambda)_{max}}(f) + l(f)\idty \quad , $$

\nind for all $f \in D(\Lambda_{max})$. \par
     Next, one can consider $\Lambda - P_n \Lambda$ instead of $\Lambda$
and prove $h \in D((P_{n-1}\Lambda)_{max})$. Repeating this process finitely
many times, one concludes that $h \in D(\Lambda_{max})$ and

$$U^* \Phi(f) U = \Phi(f) + l(f)\idty \quad , $$

\nind for all $f \in D(\Lambda_{max})$, with $U$ a suitable unitary.
As before, the boundedness of $l$ follows, and thus, again, the existence
of a $g \in H$ such that $l(f) = \sigma(g,f)$, for all $f \in H$, is assured. 
But then the equalities

$$\sigma(g,e_k) = l(e_k) = l_k = -\sigma(h,e_k) \quad , $$

\nind for all $k \in \IN$, imply that $g+h \in V$ and, thus, 
$g \in D(\Lambda)$, since $V \subset D(\Lambda)$ and $h \in D(\Lambda)$.
\hfill\qed\enddemo

\proclaim{Theorem 5.10} Let $\Lambda \in \Ls^{(n)}_{CCR}$ and $\pi$ be a
GNS-representation of a quasifree state on $\As(D(\Lambda))$. Assume that
$\pi_{\Lambda}$ and $\pi_{P_1 \Lambda}$ are irreducible (see Theorem 3.2.2).
The representations $\pi_{\Lambda}$ and $\pi$ are quasi-equivalent if and only 
if the following conditions are fulfilled: 

   (i) $\pi_{(P_0 + P_1)\Lambda}$ and $\pi$ are quasi-equivalent.

   (ii) The closure of the operator $(\idty -  P_0 - P_1) \Lambda$ is
Hilbert-Schmidt.
\endproclaim

\demo{Proof}($\Leftarrow$) Since $\pi_{P_1 \Lambda}$ is irreducible and since 
quasi-equivalence is an equivalence relation, after applying a
suitable linear transformation, it may be assumed that $P_0 \Lambda \subset 0$
and $P_1 \Lambda \subset 0$, and then Theorem 5.7 yields the quasi-equivalence
of $\pi$ and $\pi_{\Lambda}$. \par
     ($\Rightarrow$) Now let $\pi$ and $\pi_{\Lambda}$ be quasi-equivalent. 
Let $s'$ be the scalar product associated to $\pi$ and $H'$ be the completion
of $D(\Lambda)$ with respect to $s'$, as above. By Corollary 4.4, it may 
further be assumed that

$$\pi(W(f)) = e^{il(f)}e^{i\Phi(Tf)} \quad , \quad f \in D(\Lambda) \quad ,$$

\nind for a symplectic $T : D(\Lambda) \rightarrow H$ and a linear form
$l : D(\Lambda) \rightarrow \RR$. By applying a suitable coherent transformation,
one has $l \subset 0$. By Lemma 5.5, if a sequence $\{ f_n \}$ in $D(\Lambda)$ 
converges with respect to $s'$, then the strong graph limit of 
$\{ \Phi(Tf_n)\}$ exists, as does the strong graph limit of 
$\{ \Phi_{\Lambda}(f_n)\}$. From Lemma 5.4 and $P_1 \Lambda \subset 0$, one 
notes that the sequences $\{ f_n\}$ and $\{ \Lambda f_n\}$ also converge,
so it may be assumed that $D(\Lambda)$ is closed with respect to $s'$,
{\it i.e.} $T$ is closed. Theorem 5.9 entails that each coherent transformation
of $\pi_{\Lambda}$ inducing a quasi-equivalent representation is of the form

$$\Phi_{\Lambda}(f) \mapsto \Phi_{\Lambda}(f) + \sigma(f,g)\idty \quad , $$

\nind for all $f \in D(\Lambda)$ and some suitable $g \in H$. From Lemma 5.8
one concludes that such coherent transformations of $\pi$ are of the form

$$\Phi(Tf) \mapsto \Phi(Tf) + s'(f,g)\idty \quad , $$

\nind for all $f \in D(\Lambda)$ and some suitable $g \in H$. Thus, the
assumed quasi-equivalence implies that, for arbitrary $f \in D(\Lambda)$ and
arbitrary but fixed $g \in D(\Lambda)$, one has 

$$s(Tf,Tg) = s'(f,g) = \sigma(f,g') = s(f,-Jg') \quad , $$

\nind for a suitable $g' \in H$. Hence, $Tg \in D(T^*)$, for all 
$g \in D(T) = D(\Lambda)$, which implies consecutively the boundedness of
$T$ (since $T$ is closed) and then of $\Lambda$. \par
     Let $T \mid_V = U \,\vert T \mid_V \vert$ be the polar decomposition of
the restriction $T \mid_V$. By considering the quasifree quasi-equivalent 
representations
$\pi \mid \As(V)$ and $\pi_{\Lambda} \mid \As(V)$, one may conclude that
$\vert T \mid_V \vert - \idty$ is a Hilbert-Schmidt operator. Since $T$
is symplectic, one has $\sigma(f,g) = 0$, for all $f,g \in R(U)$, so that
$JR(U) \subset R(U)^{\bot}$. Let $P : H \rightarrow H$, resp. $Q : H \rightarrow H$,
be  the orthogonal projection onto $R(U)$, resp. $JR(U)$. Note that
$\vert T \mid_V \vert$ is invertible, since for any $0 \neq f \in H$ there
exists a $g \in H$ such that $\sigma(Tf,Tg) = \sigma(f,g) \neq 0$. Hence,
one can define the operator $S$ on $H$ by

$$S = U \vert T \mid_V \vert^{-1} U^* - JU \vert T \mid_V \vert U^* J +
(\idty - P - Q) \quad . $$

\nind By Lemma 4.1, $S = -JS^{-1}J$ is symplectic. Furthermore, the
operator $\vert S \vert - \idty = S - \idty$ is Hilbert-Schmidt, so we may
replace $T$ by $ST$, since quasi-equivalence is an equivalence relation.
Note that $T \mid_V = U$ is an isometry. By replacing $T$ with $WT$ for a
suitable $W$, which is unitary when viewed as an operator on $\Hs$, it
may further be assumed that $R(U) \subset V$. \par
     There exist real constants
$\lambda_{k_1 \ldots k_m}(f)$, symmetric in the indices, such that

$$\Lambda f = \underset{m,k_1,\ldots,k_m}\to{\sum} \lambda_{k_1 \ldots k_m}(f)
:x_{k_1} \cdots x_{k_m}: \quad , \quad f \in H \quad . $$

\nind As in the proof of Theorem 5.8, one applies the inverse transformation
of $\Lambda$ to $\pi_{\Lambda}$ and $\pi$. One then uses the strong graph limit
of the sequence

$$\{ \Phi_{\Lambda}(f) - \underset{m}\to{\sum}\underset{k_1,\ldots,k_m \leq l}\to{\sum} 
\lambda_{k_1 \ldots k_m}(f) :x_{k_1} \cdots x_{k_m}: \idty \}_{l \in \IN} 
\quad , $$

\nind which is $\Phi(f)$, to show that $\pi_J$ is quasi-equivalent to a
representation $\pi'$ defined by

$$\pi'(W(f)) = e^{i\Phi'(f)} \quad , f \in H \quad , $$

\nind where (note that $x(Ue_k) = \Phi(Te_k)$)

$$\Phi'(f) = \overline{\Phi(Tf) - 
\underset{m,k_1,\ldots,k_m}\to{\sum} \lambda_{k_1 \ldots k_m}(f)
:x(Ue_{k_1}) \cdots x(Ue_{k_m}):\idty} \quad . \tag{5.8} $$

     As in the proof of Theorem 5.6, one can derive Hilbert-Schmidt conditions.
In equation (5.3), one only has to replace $a(e_k)$ by $a(Ue_k)$, since the 
linear term of the annihilation operators 
$\frac{1}{\sqrt{2}}(\Phi'(e_k) + i\Phi'(Je_k))$ is, by (5.8),

$$\align
\frac{1}{\sqrt{2}}(\Phi(Te_k) + i\Phi(TJe_k)) &=
\frac{1}{\sqrt{2}}(\Phi(Ue_k) + i\Phi(TJe_k)) \\
&= a(Ue_k) + \frac{i}{\sqrt{2}} \Phi((TJ - JU)e_k) \quad .
\endalign $$

\nind In the counterpart to (5.4), one obtains the additional terms

$$
\Vert \Phi((TJ-JU)e_k)P_0 \varphi_{\epsilon} \Vert_2 \overset{(3.1.1)}\to{=}
\frac{1}{\sqrt{2}} \Vert (TJ-JU)e_k \Vert \, \Vert P_0 \varphi_{\epsilon} \Vert_2 \quad , $$

\nind and 

$$\align
\Vert \Phi((TJ-JU)e_k)\underset{l=1}\to{\overset{2n+1}\to{\sum}} P_l \varphi_{\epsilon} \Vert_2 &\leq
C \Vert (TJ-JU)e_k \Vert \Vert \underset{l=1}\to{\overset{2n+1}\to{\sum}} P_l \varphi_{\epsilon} \Vert_2 \\
&\leq C \epsilon \Vert (TJ-JU)e_k \Vert \quad .
\endalign $$

\nind One may therefore conclude that the closure of the mapping
$\Lambda - (P_0 + P_1)\Lambda$ is Hilbert-Schmidt, thus implying the 
quasi-equivalence of $\pi_{\Lambda}$ and $\pi_{(P_0 + P_1)\Lambda}$ and
hence the quasi-equivalence of $\pi$ and $\pi_{(P_0 + P_1)\Lambda}$.
\hfill\qed\enddemo

     A further immediate consequence is the following result.

\proclaim{Corollary 5.11} Let $\Lambda \in \Ls^q_{CCR} \cap \Ls^{(n)}_{CCR}$
determine an irreducible representation $\pi_{\Lambda}$ of $\As(D(\Lambda))$.
$\pi_{\Lambda}$ is quasi-equivalent to a GNS-representation of a quasifree 
state if and only if $\overline{\Lambda}$ is a Hilbert-Schmidt operator.
If $\Lambda \in \Ls^{(n)}_{CCR}$ is bounded, then $\pi_{\Lambda}$ is 
quasi-equivalent to a GNS-representation of a quasifree state if and only if 
$\overline{\Lambda - P_1 \Lambda}$ is a Hilbert-Schmidt operator.
\endproclaim

\demo{Proof}Under the given hypothesis, one has $P_1 \Lambda \subset 0$.
Thus, with Theorem 3.2.2, the stated assertions follow at once from the
previous Theorem.
\hfill\qed\enddemo

    The results in this paper can be used to provide an alternative proof to a
well-known criterion for the unitary equivalence of two pure quasifree states
\cite{33}\cite{9}. Recall that pure quasifree states are Fock.

\proclaim{Theorem 5.12} Let $\pi = \pi_J$ and $\pi' = \pi_{J'}$ be two pure
quasifree states on $\As(H_0)$, where $H_0$ is a dense subspace of $H$. Then
they are unitarily equivalent if and only if $J - J'$ is Hilbert-Schmidt
with respect to $s$, the scalar product on $H$ associated with $\pi_J$
(equivalently, Hilbert-Schmidt with respect to $s'$). 
\endproclaim

\demo{Proof} Using notation already established in Chapter IV, it follows 
easily from Lemma 5.4 that if two Fock states
$\omega_F = \omega_J$ and $\omega'_F = \omega_{J'}$ on $\As(H_0)$ are 
unitarily equivalent then the associated scalar products $s$ and 
$s'$ are equivalent - this entails $H = H'$. By Proposition 4.4, Theorems 4.5 and
5.7, the operator (employing the notation of Proposition 4.2)

$$(\vert T \vert - \idty)(\vert T \vert + \idty) = 
\vert T \vert^2 - \idty = -JK - \idty = -JJ' - \idty \quad , $$

\nind resp. $J' - J = J(-JJ' - \idty)$, is Hilbert-Schmidt with respect to
$s$, resp. $s'$.
\hfill\qed\enddemo

     As in Theorem 5.12, one can use the results of Chapters IV and V,
particularly Theorems 4.5 and 5.10 and Proposition 4.4, to give 
an alternative proof of the criteria characterizing the quasi-equivalence of 
quasifree states (see \cite{33}\cite{9}\cite{8}\cite{2}\cite{3} for 
increasingly general results), but we shall not give the details here. \par
     In \cite{11} are given necessary and sufficient conditions so that
two irreducible quadratic representations (see \cite{22}) are unitarily 
equivalent. One could generalize that result in order to obtain necessary
and sufficient conditions so that an irreducible quadratic representation
and an irreducible representation of finite degree are unitarily equivalent.
But, as these conditions are not particularly transparent, we shall not
present them here. \par
     Finally, we mention that in this paper and in the previous ones,
\cite{26}\cite{22}, the choice of the complex structure, and thus the
choice of the Fock representation, has been held fixed. It is therefore
of interest to point out that in \cite{11} the unitary equivalence of
two quadratic representations constructed from different Fock representations
is characterized. Those arguments can also be generalized to the case
of the representations of finite degree discussed in this chapter, but once
again the conditions which emerge are not particularly edifying.\par

\bigpagebreak

\heading Acknowledgements \endheading

     The research program of which this paper is a continuation would not have 
taken place without Georg Reents' initial impetus and participation. In 
addition, early versions of some of the results of this paper appeared in 
\cite{11}, a {\it Diplomarbeit} carried out under Dr. Reents' direction. We 
have also benefitted from an exchange with Prof. Paul Robinson concerning our
differing approaches to these problems.

\bigpagebreak

\heading REFERENCES \endheading

\roster
\item{}H. Araki, Hamiltonian formalism and the canonical commutation 
relations in quantum field theory, {\sl J. Math. Phys., \bf 1}, (1960), 
492--504.
\item{}H. Araki, On quasifree states of the canonical commutation relations, 
II, {\sl Publ. R.I.M.S., Kyoto Univ., \bf 7}, (1971/72), 121--152.
\item{}H. Araki and S. Yamagami, On quasi-equivalence of quasifree states
of the canonical commutation relations, {\sl Publ. R.I.M.S., Kyoto Univ.,
\bf 18}, (1982), 283--338.
\item{}F.A. Berezin, {\it The Method of Second Quantization}, Academic Press, 
New York, 1966.
\item{}O. Bratteli and D.W. Robinson, {\it Operator Algebras and Quantum 
Statistical Mechanics}, Vol. I, Springer-Verlag, New York, Heidelberg 
and Berlin, 1979.
\item{}O. Bratteli and D.W. Robinson, {\it Operator Algebras and Quantum 
Statistical Mechanics}, Vol. II, Springer-Verlag, New York, Heidelberg 
and Berlin, 1981.
\item{}J.M. Cook, The mathematics of second quantization, {\sl Trans. Amer. 
Math. Soc., \bf 74}, (1953), 222--245.
\item{}A. van Daele, Quasi-equivalence of quasi-free states on the Weyl 
algebra, {\sl Commun. Math. Phys., \bf 21}, (1971), 171--191.
\item{}A. van Daele and A. Verbeure, Unitary equivalence of Fock 
representations on the Weyl algebra, {\sl Commun. Math. Phys., \bf 20},
(1971), 268--278. 
\item{}P.A.M. Dirac, The quantum theory of the emission and absorption of 
radiation, {\sl Proc. Roy. Soc. London, \bf 114}, (1927), 243--265.
\item{}M. Florig, {\it Kanonische Transformationen von Quantenfeldern},
Diplomarbeit, University of W\"urzburg, 1995.
\item{}L. G\aa rding and A.S. Wightman, Representations of the commutation 
relations, {\sl Proc. Nat. Acad. Sci., \bf 40}, (1954), 622--626.
\item{}G.C. Hegerfeldt, G\aa rding domains and analytic vectors for quantum 
fields, {\sl J. Math. Phys., \bf 13}, (1972), 821--827.
\item{}D. Kastler, The $C^*$-algebras of a free boson field, {\sl Commun. Math.
Phys., \bf 1}, (1965), 14--48.
\item{}J.R. Klauder, J. McKenna and E.J. Woods, Direct-product representations 
of the canonical commutation relations, {\sl J. Math. Phys., \bf 7}, (1966),
822--828.
\item{}J. Manuceau, $C^*$-alg\` ebre de relations de commutation, {\sl 
Ann. Inst. Henri Poincar\'e, \bf 8}, (1968), 139--161.
\item{}J. Manuceau, F. Rocca and D. Testard, On the product form of quasi-free 
states, {\sl Commun. Math. Phys., \bf 12}, (1969), 43--57.
\item{}J. Manuceau and A. Verbeure, Quasi-free states of the CCR-algebra and
Bogoliubov transformations, {\sl Commun. Math. Phys., \bf 9}, (1968), 293--302.
\item{}J. Manuceau, M. Sirugue, D. Testard and A. Verbeure, The smallest 
$C^*$-algebra for canonical commutation relations, {\sl Commun. Math. Phys.,
\bf 32}, (1973), 231--243.
\item{}J. von Neumann, Die Eindeutigkeit der Schr\"odingerschen Operatoren,
{\sl Math. Ann., \bf 104}, (1931), 570-578.
\item{}M. Proksch, {\it Quadratische Kanonische Transformationen von 
Quantenfeldern}, Ph.D. thesis, University of W\"urzburg, 1992.
\item{}M. Proksch, G. Reents and S.J. Summers, Quadratic representations of 
the canonical commutation relations, {\sl Publ. R.I.M.S., Kyoto Univ., \bf 31},
(1995), 755--804.
\item{}M. Reed, A G\aa rding domain for quantum fields, {\sl Commun. Math. 
Phys., \bf 14}, (1969), 336--346.
\item{}M. Reed and B. Simon, {\it Methods of Modern Mathematical Physics, I.
Functional Analysis}, Academic Press, New York, 1972.  
\item{}M. Reed and B. Simon, {\it Methods of Modern Mathematical Physics, II.
Fourier Analysis, Self-Adjointness}, Academic Press, New York, 1975.  
\item{}G. Reents and S.J. Summers, Beyond coherent states: higher order 
representations, in: {\it On Klauder's Path: A Field Trip}, edited by G.G. 
Emch, G.C. Hegerfeldt and L. Streit, World Scientific, Singapore, 1994,
179--188.
\item{}D.W. Robinson, The ground state of the Bose gas, {\sl Commun. Math.
Phys., \bf 1}, (1965), 159--171. 
\item{}P.L. Robinson, Symplectic pathology, {\sl Quart. J. Math., \bf 44},
(1993), 101--107.
\item{}P.L. Robinson, Quadratic Weyl representations, {\sl Publ. R.I.M.S., 
Kyoto Univ., \bf 34}, (1998), 1--17.
\item{}P.L. Robinson, Polynomial Weyl representations, to appear in: 
{\sl Publ. R.I.M.S., Kyoto Univ.}
\item{}G. Roepstorff, Coherent photon states and spectral condition, {\sl 
Commun. Math. Phys., \bf 19}, (1970), 301--314. 
\item{}I.E. Segal, Foundations of the theory of dynamical systems of 
infinitely many degrees of freedom, I, {\sl Mat. Fys. Medd. Danske Vid. 
Selsk., \bf 31}, \# 12 (1959).
\item{}D. Shale, Linear symmetries of the free boson field, {\sl Trans. 
Amer. Math. Soc., \bf 103}, (1962), 149--167.
\item{}D. Shelupsky, Translations of the discrete Bose-Einstein operators,
{\sl J. Math. Phys., \bf 7}, (1966), 163--166. 
\item{}B. Simon, {\it The P($\Phi$)$_{2}$ Euclidean (Quantum) Field
Theory}, Princeton University Press, Princeton, 1974.
\item{}J. Slawny, On factor representations and the $C^*$-algebra of canonical
commutation relations, {\sl Commun. Math. Phys., \bf 24}, (1971), 151--170.
\item{}S.J. Summers, On the Stone-von Neumann uniqueness theorem and its
ramifications, to appear in: {\it John von Neumann and the Foundations of 
Quantum Mechanics} (tentative title), edited by M. R\'edei and M. Stoelzner.
preprint archive: mp-arc 98-720.
\endroster

\comment
\item{}H. Araki and E.J. Woods, Complete Boolean algebras of type $I$ factors,
{\sl Publ. R.I.M.S., Kyoto Univ., \bf 2}, 157-242 (1966).
\item{}R.K. Colegrave and P. Bala, Nonlinear canonical transformations, {\sl
J.Phys. A, \bf 18}, 779-791 (1985).
\item{}J. Deenen, Generators for nonlinear canonical transformations, {\sl J. 
Phys. A, \bf 24}, 3851-3858 (1991).
\item{}G.G. Emch, {\it Algebraic Methods in Statistical Mechanics and
Quantum Field Theory}, John Wiley \& Sons, New York, 1972.
\item{}R. Haag, {\it Local Quantum Physics}, Springer-Verlag, New York, 
Heidelberg and Berlin, 1992.
\item{}G.C. Hegerfeldt and J.R. Klauder, Metrics on test function spaces for 
canonical field operators, {\sl Commun. Math. Phys., \bf 16}, 329-346 (1970).
\item{}G.C. Hegerfeldt, Equivalence of basis-dependent and basis-independent
approach to canonical field operators, {\sl J. Math. Phys., \bf 12}, 167-172
(1971).
\item{}G.C. Hegerfeldt, Smoothing operators for field domains, {\sl J. Math. 
Phys., \bf 15}, 621-624 (1972).
\item{}R. Honegger and A. Rieckers, The general form of non-Fock coherent
Boson states, {\sl Publ. RIMS, Kyoto Univ., \bf 26}, 397-417 (1990).
\item{}J.R. Klauder, Exponential Hilbert space: Fock space revisited, {\sl
J. Math. Phys., \bf 11}, 609-630 (1970).
\item{}J.R. Klauder and B.-S. Skagerstam, {\it Coherent States: Applications 
in Physics and Mathematical Physics}, World Scientific, Singapore, 1985.\item{}M. Moshinsky, T.H. Seligman and K.B. Wolf, Canonical transformations 
and the radial oscillator and Coulomb problems, {\sl J. Math. Phys., \bf 13}, 
901-907 (1972).
\item{}L. Polley and U. Ritschel, Second-order phase transition in
$\lambda\phi_2 ^4$ with non-Gaussian variational approximation, {\sl Phys. 
Lett. B, \bf 221}, 44-48 (1989).
\item{}L. Polley, R.F. Streater, and G. Reents, Some covariant representations 
of massless boson fields, {\sl J. Phys. A: Math. Gen., \bf 14}, 2479-2488 
(1981).
\item{}I.E. Segal, Tensor algebras over Hilbert spaces, I, {\sl Trans. Amer. 
Math. Soc., \bf 81}, 106-134 (1956).
\item{}I.E. Segal, {\it Mathematical Problems of Relativistic Physics} 
(Providence, American Mathematical Society) 1963.
\item{}G.L. Sewell, {\it Quantum Theory of Collective Phenomena}, Oxford
University Press, Oxford, 1986.
\item{}B. Simon, {\it Functional Integration and Quantum Physics} (New York,
Academic Press) 1979.
\endcomment

\enddocument